# Dynamic drags acting on moving defects in discrete dispersive media: from dislocation to low-angle grain boundary

Soon Kim[1], Keonwook Kang[2] and Sung Youb Kim[1*]

[1]Department of Mechanical Engineering, Ulsan National Institute of Science and Technology, Ulsan 44919, South Korea

[2]Department of Mechanical Engineering, Yonsei University, Seoul 03722, South Korea

*Corresponding author

E-mail address: sykim@unist.ac.kr

Present address: EB5 801-6, 50 UNIST-gil, UNIST, Ulsan 44919, South Korea

Tel/Fax: +82-52-217-2321 / +82-52-217-4209

**Abstract**

Although continuum theory has been widely used to describe the long-range elastic behavior of dislocations, it is limited in its ability to describe mechanical behaviors that occur near dislocation cores. This limit of the continuum theory mainly stems from the discrete nature of the core region, which induces a drag force on the dislocation core during glide. Depending on external conditions, different drag mechanisms are activated that govern the dynamics of dislocations in their own way. This is revealed by the resultant speed of the dislocation. In this work, we develop a theoretical framework that generally describes the dynamic drag on dislocations and, as a result, derive a phenomenological constitutive equation. Furthermore, given that a low-angle grain boundary (LAGB) can be regarded as an array of dislocations, we extend the model to describe the mobility law of LAGBs as a function of misorientation angle. As a result, we prove that both dislocations and LAGBs follow the developed constitutive equation with the same mathematical form despite their different governing drag sources. The suggested model is also supported by molecular dynamics simulations. Therefore, this work has significance for a fundamental understanding of the dynamic drag acting on defects and facilitates a general description of various drag mechanisms.

# 1. Introduction

For a long time, dislocations have received significant scientific interest because they are elementary units that transfer material plastic deformation on an atomic level. Fundamental material properties such as strength and ductility are determined by their movement and interactions with other dislocations or defects. The overall dynamics of a dislocation are governed by its core region, although the radius of its area is extremely small, only a few atomic distances. However, there is theoretical difficulty in studying this region because of its nonlinear behaviors. This nonlinearity is closely related to the discrete nature of the dislocation core, and thus it has revealed the limitations of continuum theory, particularly when the dislocation speed approaches the shear wave speed. According to continuum theory, the dislocation cannot move faster than the transverse shear wave speed ($C_t$) because the energy density and stress field of a moving dislocation diverge at $C_t$ (Eshelby, 1956; Hirth and Lothe, 1982). In other words, the dislocation core radiates sound waves that require a supply of infinite energy to satisfy the energy balance. As a result, the shear wave speed has been regarded as a barrier that the dislocation cannot overcome (Markenscoff and Ni, 2001; Pellegrini, 2014).

However, the development of atomistic simulations has enabled researchers to simulate individual dislocations on the atomic scale. These simulation studies have shown that dislocations can move faster than $C_t$ (Koizumi et al., 2002; Wei and Peng, 2017; Peng at al., 2019), and in particular, edge dislocations can move even faster than the longitudinal shear wave speed ($C_l$) (Gumbsch and Gao, 1999; Tsuzuki et al., 2008). With these observations, the discrete lattice dynamics (DLD) theory has been suggested as a theoretical method to consider the discrete nature of dislocation cores (Atkinson and Cabrera, 1965; Swinburne and Dudarev, 2015; Verschueren et al., 2018). The DLD theory does not require the singularity that is inevitable in continuum theory. This advantage has provided a theoretical way to explain the motion of dislocation cores when their speed is close to or beyond the shear wave speed. Generally, in this speed regime, phonons are emitted from the dislocation core, and their interactions with the lattice system are significant in determining the relationship between the dislocation speed and applied stress, i.e., the mobility law. As a result of interactions between phonons and the lattice, energy dissipates from the dislocation core, which can cause drag on the moving dislocation.

There have been numerous efforts toward revealing the source of phonon emission around dislocations. First, the phonons are emitted by a scattering of elementary excitations caused by external sources. In many cases, temperature as the dominant excitation source induces oscillations in the dislocation, which radiates phonons. This behavior is called *flutter drag* (Hirth and Lothe, 1982; Amrit et al., 2018). Recently, Chen et al. (2017) showed that thermal phonons reduce the energy stored in the cores of dislocation arrays, induce the emission of secondary phonon waves, and disperse the waves around the arrays. Since this mechanism depends on the internal degrees of freedom of the dislocation, it is sensitive to temperature. Increasing temperature causes more phonons to radiate from the core, which increases the dissipation. However, the dissipation process can occur even when there are no thermal sources or external excitations. In a discrete system, when a dislocation overcomes the Peierls barrier, the changes in its core and irregular motion radiate elastic waves. This mechanism, called *radiation drag* (Kresse and Truskinovsky, 2003, 2004; Kim et al., 2016; Wang and Abeyaratne, 2018), occurs solely due to the discreteness of the system. Thus, the magnitude of the drag depends on the speed of the dislocation rather than the temperature. Especially, by considering radiation drag, the existence of a critical speed far below the shear wave speed can be proved by both DLD theory (Atkinson and Cabrera, 1965; Celli and Flytzanis, 1970) and molecular dynamics simulations (Verschueren et al., 2018). As the dislocation approaches a near-sonic speed, however, the aforementioned drags greatly increase and the resulting energy dissipation is not ignorable. This is called *relativistic effect* (Pellegrini, 2014; Gurrutxaga-Lerma, 2016; Krasnikov and Mayer, 2018; Kim et al., 2019) and it arises due to increase of the self-energy of the dislocation as its speed increases. The relativistic effect decreases the dislocation mobility with increasing the dislocation speed and causes the dislocation core spontaneously to oscillate (Kim et al., 2019). In general, the relativistic effect does not exist alone but coexists with other drag sources. Therefore, it is difficult to purely extract this effect from the other drag sources. However, in a recent work by the present authors (Kim et al., 2019), group parameters were newly defined that were proved capable of quantifying the relativistic drag force.

In addition to dislocations, there have also been studies toward understanding the dynamics of grain boundaries (GBs). Since GBs impede the motion of dislocations (Chen et al., 2012; Hughes and Hansen, 2014) or themselves act as a source to emit dislocations by accommodating external loads (Shan et al., 2004; Li et al., 2010; Quek et al., 2016), GBs have strong influences on the plastic deformation and recovery and recrystallization processes of polycrystalline materials. Thus, studying the atomic structure and dynamics of GBs has been emphasized for several decades (Sansoz and Molinary, 2005; Li and Chew, 2017; Zhang et al., 2018). Especially, since low-angle GBs (LAGBs) can alternatively be described as an array of dislocations (Lim et al., 2012; Du et al., 2015; Gu et al., 2018), the theories that have been developed to describe dislocation dynamics can be extended to LAGBs. However, the superposition of strain fields caused by the dislocations that comprise GBs causes the nonlinear behavior of the GB to become more prominent. This makes analyses of GBs more difficult and reveals the limits of linear elasticity theory for analyzing static and dynamic GB behaviors. The structural nonlinearity of GBs governs their phonon transport, which is mainly responsible for the thermal efficiency of materials (Kim et al., 2015; Yasaei et al., 2015). According to a recent work by Yasaei et al (2015), the misorientation angle of GBs is an important factor in determining the thermal resistance of graphene, which was captured through phonon scattering from GBs.

In this work, we investigate dynamic drags acting on dislocations and LAGBs and develop a general theory that is applicable to both defects by explaining their mobility laws in the frame of phonon scattering. By using DLD theory, we first derive a phenomenological constitutive equation to explain phonon drags acting on a moving dislocation. Based on this equation, we predict an unusual behavior caused by the discrete nature of its core, called stress-drop (Kim et al., 2016), which cannot be explained by continuum theory. We validate the equation by comparing it with results obtained by molecular dynamics simulations with various scattering sources. In addition to dislocations, we extend our theoretical model to LAGBs. By adopting the misorientation angle as an additional variable, the constitutive equation for LAGBs is newly derived but maintains the same form as that for dislocations. From the simulation results, we observed two unusual behaviors in addition to stress-drop while the LAGB migrates. These correspond to the curved structure of the LAGB in motion and the inverse relationship between the LAGB speed and misorientation angle. Through a systematic approach, we prove that the constitutive equation can explain these unusual behaviors and describe the mobility of the LAGB. All of these results prove that the drag caused by phonon scattering mechanisms governs the mobility of both dislocations and LAGBs in a discrete system.

## 2. Simulation methods

In this study, all atomistic simulations were performed using the Large-scale Atomic/Molecular Massively Parallel Simulator (LAMMPS) (Plimton, 1995). Through this tool, we describe the motions of various types of dislocations and LAGBs.

### 2.1. Single perfect dislocation in BCC crystals

Iron and molybdenum were chosen as media to describe the motion of a single perfect dislocation. For iron, the embedded atomic method (EAM) interatomic potential developed by Mendelev et al. (2003) was used to calculate interactions between atoms. For molybdenum, the EAM potential developed by Smirnova et al. (2013) was used. For both cases, the principal axes $x$, $y$ and $z$ were oriented along $[111]$, $[1\bar{1}0]$ and $[11\bar{2}]$ respectively. The dimensions of the simulation cells were 22.3 nm $\times$ 18.0 nm $\times$ 1.40 nm and 24.7 nm $\times$ 19.7 nm $\times$ 1.55 nm for iron and molybdenum, respectively. For both materials, we allowed the $y$ direction to be relaxed, whereas periodic conditions were applied along both the $x$ and $z$ directions, making the systems nanoplate. The potential energy of the system was minimized using the conjugate gradient method. Then, a single edge dislocation with its line along the $z$ axis was inserted by deleting the lower-half plane and applying displacement fields derived by linear elasticity theory (Hirth and Lothe, 1982) to every atom. The corresponding Burgers vector was $\mathbf{b} = 1/2[111]$. Then, we minimized the energy of the system again to determine the equilibrium dislocation structure. The equilibrium core structures of the edge dislocations in both materials are described in Fig. 1. Next, we generated an ensemble of velocities following a Gaussian distribution to produce the desired temperature. We equilibrated the system at the desired temperature using the Nosé–Hoover thermostat (Nose, 1984; Hoover, 1985) for 200 ps.

**2.2. Extended partial dislocations in FCC crystals**

In face-centered cubic (FCC) crystals, a perfect dislocation dissociates into two partial dislocations to lower the system energy (Hirth and Lothe, 1982). In this work, we chose aluminum, copper, nickel, and gold as media to insert partial dislocations. The EAM interatomic potentials developed by Mishin et al. (1999, 2001), Angelo et al. (1995), and Ackland et al. (1987) were used to calculate atomic interactions for aluminum, copper, nickel, and gold, respectively. For all materials, the $x$, $y$ and $z$ directions were oriented along $[\bar{1}10]$, $[111]$, and $[11\bar{2}]$ respectively. The dimensions of the simulation cells were 25.8 nm $\times$ 20.8 nm $\times$ 1.49 nm, 23.0 nm $\times$ 18.6 nm $\times$ 1.33 nm, 22.4 nm $\times$ 18.1 nm $\times$ 1.29 nm, and 26.0 nm $\times$ 20.9 nm $\times$ 1.50 nm for aluminum, copper, nickel, and gold, respectively. For all cases, periodic boundary conditions were applied along the $x$ and $z$ directions, and relaxation was allowed in the $y$ direction, making the systems nanoplates. The potential energy of the systems was minimized using the conjugate gradient method. Then, a single edge dislocation with its line along the z axis was inserted by deleting the lower-half plane and applying displacement fields derived by linear elasticity theory (Hirth and Lothe, 1982) to every atom. The corresponding Burgers vector was $\mathbf{b}=1/2[\bar{1}10]$. During relaxation, in contrast to the BCC case, the dislocation was divided into two partial dislocations with Burgers vectors of $\mathbf{b}_1=1/6[\bar{2}11]$ and $\mathbf{b}_2=1/6[\bar{1}2\bar{1}]$ and a stacking fault between them. The equilibrium core structures for the FCC crystals are described in Fig. 2. Next, we generated an ensemble of velocities following a Gaussian distribution to produce the desired temperature. We equilibrated the system at the desired temperature using the Nosé–Hoover thermostat (Nose, 1984; Hoover, 1985) for 200 ps.

**2.3. Low-angle grain boundaries (LAGBs)**

In order to simulate LAGBs with various misorientation angles, $\theta$, we chose a 2D triangular lattice structure as a medium rather than a 3D complex structure. The interactions between atoms were described by a Lennard–Jones (LJ) potential. The LJ potential for interatomic distance, $r$, is defined by Eq. (1).

$$V(r) = \begin{cases} 4\varepsilon \left[ \left(\frac{\sigma}{r}\right)^{12} - \left(\frac{\sigma}{r}\right)^{6} \right] & (r < r_c) \\ 0 & (r > r_c) \end{cases}, \quad (1)$$

where $\varepsilon$ and $\sigma$ are LJ parameters and $r_c$ is the cutoff radius. Here, we used $\sigma = 2.267$ Å, $\varepsilon = 0.7064\,eV$, and $r_c = 3.726$ Å as defined in a previous work by Filippova et al. (2014). Additionally, by solving $\partial V / \partial r = 0$, the equilibrium interatomic distance, $r_m$, corresponds to $2^{1/6}\sigma$, which is 2.545 Å.

We inserted the LAGBs via two steps. First, we constructed two opposite-edge dislocations separated by half of the horizontal box length in a small slab to satisfy periodic boundary conditions along the $x$ direction. Second, given that an LAGB can be described as an array of dislocations, we stacked the edge dislocations along the $y$ direction to form an LAGB. By artificially controlling the distance $D$ between two neighboring dislocations based on the relationship $\theta = b/D$, we determined the $\theta$ values for each LAGB. Although each GB has a different $\theta$ value, they are structurally identical because they consist of the same type of dislocation. After relaxation of the system using the same method described in the above sections, the equilibrium structures of LAGBs with four $\theta$ values (5.09°, 6.01°, 7.34°, and 9.42°) were obtained. To avoid difficulties arising from the system size influencing the speed of the LAGBs, we fixed the size of the systems to approximately 102 nm $\times$ 25 nm $\times$ 2.545 nm. Although there were differences in the size due to different $\theta$ values, the differences were negligible compared to the whole system. The LAGBs with four $\theta$ values are described in Fig. 3.

Furthermore, since the lattice has only one atomic layer in the $z$ direction, and the LAGBs consist of only a single type of dislocation, the gliding of dislocations within an LAGB plane does not need to be considered. Thus, under external shear stress, $\sigma_{xy}$, the motion of the LAGBs is confined to glide along the $x$ direction on the $xz$ plane.

### 2.4. Application of constant shear stress

For both the edge dislocations and LAGBs, a constant shear force was applied to every atom in the top and bottom free surfaces (or edges) along the $x$ axis in opposite directions. Each atomic position was updated every 1 fs using the Nosé–Hoover thermostat (Nose, 1984; Hoover, 1985) at the desired temperature. As a result, both the dislocations and LAGBs moved toward the $x$ direction.

### 2.5. Measurement of defect speeds and actual stress

We measured the speeds of the dislocations and LAGBs by tracing their core positions. The core atoms are characterized by a unique common neighbor analysis pattern compared to atoms in a perfect region. Then, we defined the core position as a single point by averaging and recording the positions of core atoms every 1 ps. Finally, we determined the velocity of each defect by measuring the slope of the relationship between simulation time and distance that the core moved after the system reached an equilibrium state.

According to previous studies (Kim et al.,2016; Cho et al., 2017; Kim et al., 2019), average stress around the dislocation core is different from the externally applied stress while the dislocation moves. Therefore, we measured the average stress and defined it as $\sigma_{act}$, or *actual stress* in this study. For dislocations, we measured it by averaging stresses of all atoms within a circular region that includes the core at its center. Here, we chose a radius of the circular region approximately as $10b$. For LAGBs, we measured $\sigma_{act}$ by averaging stresses of all internal atoms within the system since it is difficult to calculate $\sigma_{act}$ when a distance between two opposite LAGBs is small.

## 3. Simulation results: stress-drop and the effect of misorientation angle

### 3.1. Dislocations

Depending on whether the magnitude of the externally applied stress, $\sigma_{app}$, is larger than the Peierls stress $\sigma_P$, two different phenomena can occur. For an edge dislocation in an iron nanoplate, when $\sigma_{app}$ was smaller than $\sigma_P$ such that the dislocation does not move, the actual stress around the dislocation $\sigma_{act}$ was equal to $\sigma_{app}$, as shown in Fig. 4(a). On the other hand, when $\sigma_{app}$ was larger than $\sigma_P$, which causes the dislocation to move, $\sigma_{act}$ became smaller than $\sigma_{app}$ while the dislocation was in motion. This is described in Fig. 4(b).

We defined this behavior as *stress-drop* in our previous study (Kim et al., 2016). The same behavior was also observed in other crystals whenever the dislocation moved as shown in Fig. 4(c) to (g). For the dislocations in molybdenum and FCC crystals, however, since they have extremely low $\sigma_P$ below 20 MPa, the response of the system under $\sigma_{app}$ smaller than $\sigma_P$ is not shown due to accuracy issues with the simulation.

**3.2. LAGBs**

In order to move LAGBs, a much higher stress than $\sigma_P$ for a single dislocation should be applied to the free edges of the system. This is because additional energy is required for each dislocation that comprises the LAGB to overcome the energy barrier originating from interactions with other dislocations. The required energy increases as $\theta$ increases because of the decreased distances between adjacent dislocations.

While the LAGBs were in motion, as in the motion of dislocations, the stress-drop phenomenon was also observed as described in Fig. 5 for various $\theta$. Fig. 5(a), (c), (e), and (g) show $\sigma_{act}$ when $\sigma_{app}$ was lower than the critical stress to migrate the corresponding LAGB, and Fig. 5(b), (d), (f), and (h) show $\sigma_{act}$ when $\sigma_{app}$ was higher than the critical stress. During LAGB motion, $\sigma_{act}$ was lower than $\sigma_{app}$ and remained steady for a while in a quasi-equilibrium state. It finally converged to $\sigma_{app}$ when two opposite LAGBs approached and annihilated each other under a continuously applied load.

In addition to stress-drop, two additional unusual behaviors were observed during the motion of the LAGBs. First, the speed of the LAGBs $v_{LAGB}$ decreased as $\theta$ increased under the same applied stress as described in Fig. 6. This is opposite of the result obtained by continuum dislocation theory. According to the previous continuum theory (Sutton and Balluffi 1995; Winning et al., 2010), the stress-driven LAGB motion follows

$$v_{LAGB} = M_{LAGB}\sigma\theta, \qquad (2)$$

where $M_{LAGB}$ is the mobility of the LAGB. In Fig. 6, we did not record the speed of LAGBs with misorientation angles lower than 7.34° moving under $\sigma_{app} > 6.5\,\text{GPa}$ because failures within the LAGB structures occurred during their motion. Since a decreased $\theta$ means an increase in distance between neighboring dislocations that comprise the LAGB, these failures might occur because the applied stress was enough to overcome interactions among them. Second, although the structure of an LAGB should be a straight line once it reaches an equilibrium state, as proved in Appendix A based on linear elasticity theory, the simulations showed that the LAGB was rather curved for all $\theta$ values as shown in Fig. 7. Furthermore, the magnitude of curvature decreased as $\theta$ increased under the same load.

## 4. Theoretical models for defect motion

To explain the unusual behaviors reported in Section 3 and further describe the dynamic defect behaviors, we must consider the discrete nature of the dislocation core and the resulting phonon scattering, which cannot be fully explained by linear elasticity theory. For this, we have developed a theoretical model based on DLD theory for the simplest type of defects—non-oscillating dislocations (Kim et al., 2016). Here, we reproduce the theoretical model for completeness and extend it to cases where the dislocation core oscillates and finally to LAGBs.

## 4.1. Non-oscillating dislocation

From an atomic point of view, the motion of a dislocation consists of repeated breaking and reforming of atomic bonds in its core. Owing to the anharmonicity of the strain field around the core, the elastic waves emitted when an atomic bond is broken are scattered. This scattering process shifts frequencies of the radiated waves and dissipates energy around the dislocation core supplied by $\sigma_{app}$. As a result, the average stress around the moving dislocation decreases to $\sigma_{act}$. This is the dominant drag mechanism affecting dislocation motion when thermal fluctuation is ignorable (Ohashi, 1968; Kim et al., 2016).

For simplicity, let us first consider a 2D lattice and assume that the lattice consists of atoms—each with a mass of $M$—connected by springs of stiffness $K$. Our lattice system is graphically described in Fig. 8(a). Assume that the core moves forward at a constant speed, $v$, under $\sigma_{app}$. During its motion, phonon scattering occurs around the core, which drags it backward. Thus, the core advances a distance less than expected. This is described in Fig. 8(b), 8(c), and 8(d) with a detailed explanation. If we define the distance that the core actually advances as $x_{act}$, the expected distance that the core advances due to $\sigma_{app}$ as $x_{app}$, and the receded distance by the scattering due to the anharmonic strain field as $x_{bond}$, they are mathematically related by Eq. (3). Physically, $x_{bond}$ represents a pullback distance because of the energy loss by the scattering.

$$x_{act} = x_{app} - x_{bond},$$

or

$$\frac{x_{act}}{x_{app}} = \frac{\sigma_{act}}{\sigma_{app}} = 1 - \frac{x_{bond}}{x_{app}}. \tag{3}$$

We can validate Eq. (3) by giving two examples. First, when the core does not move, the atomic bonds in the core are not broken, and thus no waves are emitted. This leads $x_{bond}$ to be 0 and $\sigma_{act} = \sigma_{app}$. On the other hand, when the core moves, the atomic bonds are broken, and the resulting emitted waves are scattered by the anharmonic strain field. Therefore, $x_{bond}$ is positive and $\sigma_{act}$ becomes smaller than $\sigma_{app}$. These results coincide with the stress-drop phenomenon described in Fig 4.

By using the (retarded) lattice Green's function, $G(\mathbf{r} - \mathbf{r_0}, w)$, a relationship between $x_{bond}$ and $x_{app}$ can be obtained. $G(\mathbf{r} - \mathbf{r_0}, w)$ is defined by Eq. (4) and is derived in detail in Appendix B.

$$G(\mathbf{r} - \mathbf{r_0}, w) = \frac{b^2}{4\pi^2 M} \int_{1st\,BZ} d\mathbf{k} \frac{e^{i\mathbf{k}\cdot\mathbf{r}}}{w^2(\mathbf{k}) - w^2 \pm i\varepsilon} \quad (\varepsilon \to +0). \tag{4}$$

Here, $w(\mathbf{k}) = w_0[\sin^2(k_x b/2) + \sin^2(k_y b/2)]^{1/2}$, where $w_0 = 2(K/M)^{1/2}$, is a dispersion relation of phonons in the 2D square lattice system, and $\vec{r}$ is the position of an atom in the discrete lattice corresponding to $(mb, nb)$, where both $m$ and $n$ are integer multiples of 1/2. The sign of $i\varepsilon$ in Eq. (4) is determined to satisfy causality conditions. That is, if $\text{Re}(w) < 0$, $+i\varepsilon$ is chosen, but if $\text{Re}(w) > 0$, $-i\varepsilon$ is chosen.

The elastic wave emitted by atomic-bond breaking at $t=0$ in the core, $f(t)$, is described by using a Heaviside step function. By applying a Fourier transformation, the wave can be expressed as

$$f(t) = \begin{cases} 0 & (t<0) \\ Kx_{app} = \dfrac{1}{4}Mw_0^2 x_{app} & (t>0) \end{cases}$$

$$= \frac{1}{2\pi}\int_{-\infty}^{\infty} F(w)e^{-iwt}dw = -\frac{Mw_0^2 x_{app}}{8\pi i}\int_{-\infty}^{\infty}\frac{e^{-iwt}}{w+i\varepsilon}dw \quad (\varepsilon \to +0), \tag{5}$$

where $F(w) = -Mw_0^2 x_{app}/(4i(w+i\varepsilon))$. If we assume that $\mathbf{r_0} = \mathbf{0}$ is the position where the wave is emitted, the displacement of the atom induced by the scattering of the wave is derived as

$$u(\mathbf{r},t) = \frac{1}{2\pi}\int_{-\infty}^{\infty} G(\mathbf{r},w)F(w)e^{-iwt}dw$$

$$= -\frac{w_0^2 b^2 x_{app}}{32\pi^3 i}\int_{1st\,BZ} e^{i\mathbf{k}\cdot\mathbf{r}}d\vec{\mathbf{k}}\int_{-\infty}^{\infty}\frac{e^{-iwt}}{[w^2(\mathbf{k})-w^2 \pm i\varepsilon](w+i\varepsilon)}dw. \tag{6}$$

By using residue theorem, Eq. (6) is transformed to Eq. (7).

$$u(\mathbf{r},t) = \frac{w_0^2 b^2 x_{app}}{32\pi^3 i}\int_{1st\,BZ} e^{i\mathbf{k}\cdot\mathbf{r}}d\mathbf{k}\left[\begin{array}{l}\oint_{C_1}\dfrac{e^{-iwt}}{[w-(w(\mathbf{k})+i\varepsilon)][w+(w(\mathbf{k})+i\varepsilon)](w+i\varepsilon)}dw \\ +\oint_{C_2}\dfrac{e^{-iwt}}{[w-(w(\mathbf{k})-i\varepsilon)][w+(w(\mathbf{k})-i\varepsilon)](w+i\varepsilon)}dw\end{array}\right]$$

$$= \frac{w_0^2 b^2 x_{app}}{32\pi^3 i}\int_{1st\,BZ} e^{i\mathbf{k}\cdot\mathbf{r}}d\mathbf{k}\left[\oint_{C_1} A_1(w)dw + \oint_{C_2} A_2(w)dw\right], \tag{7}$$

$$= \frac{w_0^2 b^2 x_{app}}{32\pi^3 i}\int_{1st\,BZ} e^{i\mathbf{k}\cdot\mathbf{r}}d\mathbf{k}\cdot 2\pi i\left[\operatorname*{Res}_{w=-w(\mathbf{k})-i\varepsilon} A_1(w) + \operatorname*{Res}_{w=-i\varepsilon} A_1(w) + \operatorname*{Res}_{w=w(\mathbf{k})-i\varepsilon} A_2(w)\right]$$

where $C_1$ and $C_2$ are contours defined in Fig. 9. Then, Eq. (7) can be simplified to Eq. (8).

$$u(\mathbf{r},t) = \frac{w_0^2 b^2 x_{app}}{16\pi^2}\int_{1st\,BZ}\frac{1-\cos(w(\mathbf{k})t)}{w^2(\mathbf{k})}e^{i\mathbf{k}\cdot\mathbf{r}}d\mathbf{k}$$

$$= \frac{w_0^2 b^2 x_{app}}{4\pi^2}\int_0^{\pi/b}\int_0^{\pi/b}\frac{\cos(mk_x b)\cos(nk_y b)}{w^2(k_x,k_y)}[1-\cos(w(k_x,k_y)t)]dk_x dk_y. \tag{8}$$

As shown in Fig. 8(b), the receded distance of the core caused by phonon scattering is derived as the difference between atoms that are located above and below the slip plane when the core reaches the neighboring Peierls valley, or $t = b/2v$. Therefore,

$$x_{bond} = u\left(\left(\frac{b}{2}, \frac{b}{2}\right), \frac{b}{2v}\right) - u\left(\left(\frac{b}{2}, -\frac{b}{2}\right), \frac{b}{2v}\right). \tag{9}$$

The existence of a dislocation imposes the boundary condition $u((mb,-nb),t) = -u((m,n),t)$. As a result, Eq. (9) becomes

$$\begin{aligned}
x_{bond} &= 2u\left(\left(\frac{b}{2}, \frac{b}{2}\right), \frac{b}{2v}\right) \\
&= \frac{w_0^2 b^2 x_{app}}{2\pi^2} \int_0^{\pi/b} \int_0^{\pi/b} \frac{\cos(k_x b/2)\cos(k_y b/2)}{w^2(k_x, k_y)} \left[1 - \cos\left(\frac{w(k_x, k_y)b}{2v}\right)\right] dk_x dk_y \\
&\approx \frac{w_0^2 b^2 x_{app}}{16\pi^2} \frac{1}{v^2} \left(\frac{2}{b}\right)^2 \int_0^{\pi/2} \int_0^{\pi/2} \cos x \cos y \, dx dy = \frac{w_0^2 b^2 x_{app}}{4\pi^2 v^2}.
\end{aligned}$$

(10)

If Eq. (10) is inserted into Eq. (3), $\sigma_{act}$ and $\sigma_{app}$ can be simply related by Eq. (11). Therefore, as long as the dislocation is in motion, $\sigma_{act}$ is always lower than $\sigma_{app}$. This was systemically proved in our previous study (Kim et al., 2016).

$$\sigma_{act} = \left(1 - \frac{w_0^2 b^2}{4\pi^2 v^2}\right)\sigma_{app}. \tag{11}$$

Furthermore, we can extract physical meaning from Eq. (11). The difference between $\sigma_{act}$ and $\sigma_{app}$ is caused by the second term on the right-hand side of Eq. (11). This term is directly proportional to the square of $t_P$, where $t_P = b/2v$, which is the time required for the core to arrive at a neighboring Peierls valley. Physically, as the dislocation speed increases, less time is required to reach to the neighboring valley, and thus the system does not have enough time for scattering to occur. Therefore, $\sigma_{act}$ converges to $\sigma_{app}$.

**4.2. Oscillating dislocation**

In general, a moving dislocation is accompanied by oscillation. There are two types of oscillation sources. First, the dislocation can be forced to oscillate due to external sources such as thermal fluctuations or elastic interactions with other defects. On the other hand, the dislocation can also oscillate spontaneously without external sources when its speed is close to that of the transverse shear wave ($C_t$). This is called the relativistic effect and it still exists even when the temperature is extremely low (Kim et al., 2019). When the relativistic drag force is large, the change in dislocation speed with increasing external stress becomes smaller such that the mobility of the dislocation core decreases. In the majority of cases, oscillations exist not as separated, single units but instead as multiple sources coexisting.

Although the sources of oscillations are different, they have the same effect of dissipating energy from the core to its surroundings and causing drag effect on the motion of the dislocation. Therefore, in this work, we simply describe a displacement of the oscillating core atom with frequency $\Omega$ as $x(t) = x_{app} \sin(\Omega t)$. If the oscillation effect is considered in addition to the breaking of atomics bonds, Eq. (3) becomes Eq. (12).

$$x_{act} = x_{app} - x_{bond} - x_{osc},$$

or

$$\frac{x_{act}}{x_{app}} = \frac{\sigma_{act}}{\sigma_{app}} = 1 - \frac{x_{bond}}{x_{app}} - \frac{x_{osc}}{x_{app}}. \qquad (12)$$

To relate $x_{osc}$ and $x_{app}$, the force originating from the oscillation, $f_{osc}$, is first described. Since our goal is to simply capture the effect of the oscillation rather than its cause, we use the relationship $f_{osc} = D\dot{x}(t)$, where $D$ is the drag coefficient of the oscillation. Via a Fourier transform, $f_{osc}$ can be alternatively described by Eq. (13).

$$f_{osc}(t) = \frac{1}{2\pi} \int_{-\infty}^{\infty} F_{osc}(w) e^{-iwt} dw$$
$$= \frac{D\Omega x_{app}}{2} \int_{-\infty}^{\infty} [\delta(-w-\Omega) + \delta(-w+\Omega)] e^{-iwt} dw, \tag{13}$$

where $F_{osc}(w) = \pi D\Omega_{osc} x_{app} [\delta(-w-\Omega) + \delta(-w+\Omega)]$.

As the dislocation oscillates, elastic waves are emitted and they are scattered around the dislocation core by anharmonic strain field. The scattered displacement of the core atom induced by the oscillation, $u_{osc}(\mathbf{r},t)$, can be derived as

$$u_{osc}(\mathbf{r},t) = \frac{1}{2\pi} \int_{-\infty}^{\infty} G(\mathbf{r},w) F_{osc}(w) e^{-iwt} dw$$
$$= \frac{D\Omega b^2 x_{app}}{8\pi^2 M} \int_{1st\,BZ} e^{i\mathbf{k}\cdot\mathbf{r}} d\vec{\mathbf{k}} \int_{-\infty}^{\infty} \frac{\delta(-w-\Omega) + \delta(-w-\Omega)}{w^2(\mathbf{k}) - w^2} e^{-iwt} dw. \tag{14}$$

Eq. (14) can be simplified to Eq. (15) as

$$u_{osc}(\mathbf{r},t) = \frac{D x_{app}}{\pi^2 M\Omega} J(\mathbf{r}) \cos(\Omega t), \tag{15}$$

where

$$J(\mathbf{r}) = \int_0^\pi \int_0^\pi \frac{\cos(mx)\cos(ny)}{\frac{w^2(x,y)}{\Omega^2} - 1} dxdy. \tag{16}$$

The receded distance of the core caused by the oscillation, $x_{osc}$, can now be derived as Eq. (17). Here, we assume that the core speed is much higher than that of the oscillation given that the externally applied stress is larger than the drag force.

$$\begin{aligned}x_{osc} &= u_{osc}\left(\left(\frac{b}{2},\frac{b}{2}\right),\frac{b}{2v}\right) - u_{osc}\left(\left(\frac{b}{2},-\frac{b}{2}\right),\frac{b}{2v}\right) \\ &= 2u_{osc}\left(\left(\frac{b}{2},\frac{b}{2}\right),\frac{b}{2v}\right) \\ &= \frac{2DJ_{dis}x_{app}}{\pi^2 M\Omega}\cos\left(\frac{\Omega b}{2v}\right) \approx \frac{2DJ_{dis}x_{app}}{\pi^2 M\Omega}\left(1-\frac{\Omega^2 b^2}{8v^2}\right).\end{aligned} \quad (17)$$

In Eq. (17), $J_{dis}$ is defined as

$$J_{dis} \equiv \int_0^\pi \int_0^\pi \frac{\cos\left(\frac{x}{2}\right)\cos\left(\frac{y}{2}\right)}{\left(\frac{w_0}{\Omega}\right)^2 \left(\sin^2\left(\frac{x}{2}\right)+\sin^2\left(\frac{y}{2}\right)\right)-1} dxdy. \quad (18)$$

By inserting Eqs. (10) and (17) into Eq. (12), $\sigma_{act}$ and $\sigma_{app}$ can be simply related by Eq. (19).

$$\sigma_{act} = \left(1 - C_1^{dis} + \frac{w_0^2 b^2}{4\pi^2 v^2}(C_2^{dis}-1)\right)\sigma_{app}. \quad (19)$$

Here, $C_1^{dis}$ and $C_2^{dis}$ are dimensionless parameters that are defined as

$$\begin{aligned}C_1^{dis} &= \frac{2DJ_{dis}}{\pi^2 M\Omega}, \\ C_2^{dis} &= \frac{DJ_{dis}\Omega}{Mw_0^2}.\end{aligned} \quad (20)$$

If we insert Eq. (19) into the relationship $v = M_{dis}\sigma_{act}$, where $M_{dis}$ is the mobility of the dislocation core, Eq. (21) is derived as

$$v^3 - M_{dis}\sigma_{app}(1-C_1^{dis})v^2 - M_{dis}\sigma_{app}\frac{w_0^2 b^2}{4\pi^2}(C_2^{dis}-1) = 0.\tag{21}$$

### 4.3. LAGB

For $\theta \leq 10°$, an LAGB can be regarded as an array of dislocations (Read and Shockley, 1950; Hull and Bacon, 2011). For a LAGB, unlike a single dislocation, interactions among dislocations must also be considered to analyze its dynamics. The motion of the LAGB consists of two steps, before and after the dynamic equilibrium states.

Initially, the LAGB begins to move via the motion of dislocations closest to the free edge of the system. Internal dislocations move in turn, and finally, the dislocation located at the center of the system moves as described in Fig. 10(b). This is because the stress applied to the edge propagates in the form of a wave toward the inside of the system. Thus, the different start times of the motion of each dislocation induce the curvature of the LAGB. This leads the horizontal distance, $x_h$, between two neighboring dislocations to be equal to $vb/C_t\theta$. From Fig. 6, $v$ decreases as $\theta$ increases. As a result, the increasing misorientation angle causes $x_h$ to decrease such that the curvature of the LAGB decreases. This is identical to the simulation results described in Fig. 7.

Once the LAGB is in steady motion, however, it maintains the curved shape despite its high potential energy as described in Fig. 10(c). If an imaginary stress, $\sigma_{img}$, acts on the system to cause it to be in equilibrium with the curved LAGB, the force equilibrium equation should satisfy Eq. (22) for each constituent dislocation. The forces that satisfy Eq. (22) are denoted by arrows in Fig. 11. Based on the figure, $\sigma_{img}$ always points toward the direction opposite the motion of the LAGB. Therefore, it is expected that $\sigma_{img}$ drags the LAGB motion.

$$\sigma_{int}^{j} + \sigma_{app} + \sigma_{img} = 0 \quad (\text{for } j = 1, 2, \cdots N), \tag{22}$$

where

$$\sigma_{int}^{j} = \frac{Gb}{2\pi(1-v)} \sum_{\substack{i=1 \\ (i \neq j)}}^{N} \frac{(x_j - x_i)[(x_j - x_i)^2 - (y_j - y_i)^2]}{[(x_j - x_i)^2 + (y_j - y_i)^2]^2}.$$

When the LAGB is in motion, it does not rigidly translate but is instead accompanied by transverse oscillations because its constituent dislocations are interacting with each other. Although the exact calculation of the oscillation amplitude of each dislocation is quite difficult, we roughly describe it in the frame of the discrete lattice model by simply assuming that each dislocation is connected by a spring that is elongated by the external stress. Then, the oscillation amplitude, $x_{osc}$, can be expressed as $x_{osc} = \alpha x_h = \alpha v b / C_t \theta$, where $0 < \alpha < 1$. Since the speed of the dislocation is roughly proportional to the applied stress, we can alternatively express the oscillation amplitude here as $x_{osc} = \Psi x_{app} / \theta$, where $\Psi$ is a function of applied stress. As a result, the elongation as a function of time becomes $x(t) = (\Psi x_{app} / \theta) \sin(\Omega_{LAGB} t)$, where $\Omega_{LAGB}$ is the LAGB oscillation frequency. This elongation induces the oscillation of each dislocation, which becomes a source of drag. This drag causes the LAGB to maintain its curved structure. As mentioned in Section 4.2, our goal is not an exact derivation of the force acting on the LAGB but a simple description of its resultant behavior, and so further mathematical derivation is identical to that used in Section 4.2. As a result, the actual internal stress is related to the external stress by

$$\sigma_{act} = \left(1 - \frac{C_1^{LAGB}}{\theta} + \frac{w_0^2 b^2}{4\pi^2 v^2}\left(\frac{C_2^{LAGB}}{\theta} - 1\right)\right)\sigma_{app}, \tag{23}$$

where

$$C_1^{LAGB} = \frac{2D\Psi J_{LAGB}}{\pi^2 M \Omega_{LAGB}},$$
$$C_2^{LAGB} = \frac{D\Psi J_{LAGB} \Omega_{LAGB}}{M w_0^2}. \tag{24}$$

In Eq. (24), $J_{LAGB}$ is defined as

$$J_{LAGB} \equiv \int_0^\pi \int_0^\pi \frac{\cos\left(\frac{x}{2}\right)\cos\left(\frac{y}{2}\right)}{\left(\frac{w_0}{\Omega_{LAGB}}\right)^2 \left(\sin^2\left(\frac{x}{2}\right) + \sin^2\left(\frac{y}{2}\right)\right) - 1} dxdy. \tag{25}$$

Therefore, the constitutive equation of motion of the LAGB becomes

$$v^3 - M_{dis}\sigma_{app}\left(1 - \frac{C_1^{LAGB}}{\theta}\right)v^2 - M_{dis}\sigma_{app}\frac{w_0^2 b^2}{4\pi^2}\left(\frac{C_2^{LAGB}}{\theta} - 1\right) = 0. \tag{26}$$

## 5. Discussion of the theoretical model

In Section 4, we proved that oscillating dislocations generally follow Eq. (21) and can be extended to LAGBs according to Eq. (26), although the exact values of $C_1^{dis}$ and $C_2^{dis}$ (or $C_1^{LAGB}$ and $C_2^{LAGB}$) depend on the source of oscillations. In this section, we discuss the significance of both parameters to understand the drag effect acting on gliding dislocations and LAGBs. To determine $C_1^{dis}$ and $C_2^{dis}$ from Eq. (20), it is required to first calculate $J_{dis}$.

Thus, we begin our discussion by extracting the physical meaning of $J_{dis}$ through a lattice dynamics analysis.

Here, we adopt a method suggested by Lifshitz and Kosevich (1966) in which we assume that the dislocation core is an isolated perturbation (Montroll and Potts, 1955; Maradudin et al., 1958). In general, a system including this perturbation follows Eq. (27).

$$\frac{\partial^2 \mathbf{u}}{\partial t^2} + \mathbf{L}\mathbf{u} + \Lambda(\mathbf{r}_p)\mathbf{u} = \mathbf{0}, \tag{27}$$

where $\mathbf{L}$ is a diagonal operator that represents the square of the frequency band of the system, $\Lambda(\mathbf{r}_p)$ is a local perturbation matrix, and $\mathbf{r}_p$ is the position of the perturbation. Here, we assume that $\Lambda(\mathbf{r}_p)$ only affects the immediate neighbors of the perturbation point. Under the plane wave assumption, we insert $\mathbf{u}(\mathbf{r},t) = \mathbf{x}(\mathbf{r})e^{i\Omega t}$ into Eq. (27). If we define $f(\mathbf{r})$ to be decreasing rate of the local perturbation away from its center, Eq. (28) can be derived (Lifshitz and Kosevich, 1966).

$$\Omega^2 x(\mathbf{r}) - \sum_{\mathbf{r}'} L(\mathbf{r} - \mathbf{r}')x(\mathbf{r}') = \Lambda(\Omega^2)f(\mathbf{r})\sum_{r'} f(\mathbf{r}')x(\mathbf{r}'). \tag{28}$$

Here, we assume that $\mathbf{r_p} = \mathbf{0}$ and $f(\mathbf{r})$ should satisfy the normalization condition $\sum_{\mathbf{r}} f(\mathbf{r}) = 1$. To solve Eq. (28), we use the Green's function method. The Green' function is determined by the following equation.

$$\sum_{\mathbf{r}'} L(\mathbf{r} - \mathbf{r}')G(\Omega;\mathbf{r}') - \Omega^2 G(\Omega;\mathbf{r}) = \delta(\mathbf{r}), \tag{29}$$

where $\delta(\mathbf{r})$ is a Dirac delta function. Applying a Fourier transform to both sides of Eq. (29) yields

$$G(\Omega;r) = \sum_{\mathbf{k}} G(\Omega;\mathbf{k})e^{i\mathbf{k}\cdot\mathbf{r}} = \sum_{\mathbf{k}} \frac{1}{w_0^2(\mathbf{k}) - \Omega^2} e^{i\mathbf{k}\cdot\mathbf{r}}, \tag{30}$$

where $w_0^2(\mathbf{k}) = \sum_{\mathbf{r}} L(\mathbf{r})e^{-i\mathbf{k}\cdot\mathbf{r}}$. Now, we apply a Fourier transform to both sides of Eq. (28) to give

$$\{w_0^2(\mathbf{k}) - \Omega^2\}x_{\mathbf{k}} = -\Lambda(\Omega^2)\sum_{\mathbf{r}}\sum_{\mathbf{r'}} f(\mathbf{r})f(\mathbf{r'})x(\mathbf{r'})e^{-i\mathbf{k}\cdot\mathbf{r}}. \tag{31}$$

Eq. (31) can be transformed into

$$\begin{aligned} x(\mathbf{r}) &= -\Lambda(\Omega^2)\left[\sum_{\mathbf{r'}} f(\mathbf{r'})\right]\left[\sum_{\mathbf{k}} G(\Omega;\mathbf{k})e^{i\mathbf{k}\cdot(\mathbf{r}-\mathbf{r'})}\right]\left[\sum_{\mathbf{r_0}} f(\mathbf{r_0})x(\mathbf{r_0})\right], \\ &= -\Lambda(\Omega^2)\sum_{\mathbf{r'}} G(\Omega;\mathbf{r}-\mathbf{r'})f(\mathbf{r'})\tilde{x} \end{aligned} \tag{32}$$

where $\tilde{x} = \sum_{\mathbf{r}} f(\mathbf{r})x(\mathbf{r})$. We multiply both sides of Eq. (32) by $\sum_{\mathbf{r}} f(\mathbf{r})$ to give

$$1 + \Lambda(\Omega^2)\sum_{\mathbf{k}} \frac{|a(\mathbf{k})|^2}{w_0^2(\mathbf{k}) - \Omega^2} = 0, \tag{33}$$

where $a(\mathbf{k}) = \sum_{\mathbf{r}} f(\mathbf{r})e^{-i\mathbf{k}\cdot\mathbf{r}}$. If we define $D(\Omega^2)$ as

$$D(\Omega^2) \equiv \sum_{\mathbf{k}} \frac{|a(\mathbf{k})|^2}{w_0^2(\mathbf{k}) - \Omega^2} = \frac{1}{(2\pi)^2}\int \frac{|a(\mathbf{k})|^2}{w_0^2(\mathbf{k}) - \Omega^2} d\mathbf{k}, \tag{34}$$

then Eq. (33) can be rewritten as $\Lambda(\Omega^2)D(\Omega^2) = -1$. As a result, this relationship determines the permissible vibrational frequencies near the local perturbation (Lifshitz and Kosevich, 1966). From Eq. (34), when $|a(\mathbf{k})|^2 = \cos k_x b \cos k_y b$, $D(\Omega^2)$ and $J_{dis}$ can be related by

$$J_{dis} = (2\pi\Omega)^2 D(\Omega^2). \tag{35}$$

Since $|a(\mathbf{k})|^2$ is proportional to the decreasing rate of the perturbation, $J_{dis}$ quantifies the localization of local vibrations around the dislocation frequency, $\Omega$, in reciprocal space. Now, we extract the physical meaning of $J_{dis}$ in *real* space through a further mathematical process. Since $\Lambda(\Omega^2) = -(2\pi\Omega)^2 / J_{dis}$, Eq. (32) becomes

$$x(\mathbf{r}) = \frac{(2\pi\Omega)^2}{J_{dis}} \tilde{x} \sum_{\mathbf{k}} \frac{a(\mathbf{k})e^{i\mathbf{k}\cdot\mathbf{r}}}{w_0^2(\mathbf{k}) - \Omega^2}. \tag{36}$$

At a large distance from the perturbation source, $a(\mathbf{k}) \approx 1$ because $f(\mathbf{r}) \approx \delta(\mathbf{r})$, and thus small wavenumbers become important. For small wavenumbers, it can be assumed that $w_0^2(\mathbf{k}) \approx w_{0,min}^2 + \beta^2 k^2$ for an isotropic model where $\beta^2$ is a constant with a magnitude on the order of $(w_{0,max}^2 - w_{0,min}^2)$. Therefore, at a large distance from the defect,

$$\begin{aligned}\frac{x(\mathbf{r})}{\tilde{x}} &= \frac{\Omega^2}{J_{dis}} \int \frac{e^{i\mathbf{k}\cdot\mathbf{r}}}{w_0^2(\mathbf{k}) - \Omega^2} d\mathbf{k} \approx \frac{\Omega^2}{\beta^2 J_{dis}} \int \frac{e^{i\mathbf{k}\cdot\mathbf{r}}}{k^2 + (w_{0,min}^2 - \Omega^2)/\beta^2} d\mathbf{k} \\ &= \frac{\pi l \Omega^2}{\beta^2 J_{dis}} e^{-r/l}\end{aligned}, \tag{37}$$

where $l = \sqrt{\beta^2/(w_{0,min}^2 - \Omega^2)}$. According to Eq. (37), $x(r)$ decreases as $J_{dis}$ increases. In other words, as the frequencies become more localized around the frequency of dislocation, the atomic amplitude far from the dislocation core decreases. This implies that $J_{dis}$ also represents the *structural compactness* of the dislocation core in real space. As a result, a large $J_{dis}$ of the dislocation core indicates a condensed structure, whereas a small $J_{dis}$ indicates a widely extended core.

From the meaning of $J_{dis}$, we can investigate the influence of group parameters $C_1^{dis}$

and $C_2^{dis}$ on the dislocation motion. According to Eqs. (12) and (19), the magnitude of the oscillation drag force relative to the externally applied force can be simply determined by $C_1^{dis}$ and $C_2^{dis}$ as shown in Eq. (38).

$$\frac{F_{osc}}{F_{app}} = \frac{x_{osc}}{x_{app}} = C_1^{dis} - \left(\frac{w_0 b}{2\pi v}\right)^2 C_2^{dis}. \tag{38}$$

Since both $C_1^{dis}$ and $C_2^{dis}$ are proportional to $J_{dis}$ per Eq. (20), it is expected that a more compact core structure will lead to an increased oscillation drag force. This may lead to the fact that a compact dislocation requires a higher Peierls stress and is less mobile than an extended dislocation under the same external conditions if other conditions are the same (Mills and Stadelmann, 1989; Srinivasan et al., 2005). According to Eq. (38), the oscillation drag force increases as $C_1^{dis}$ increases and $C_2^{dis}$ decreases. As a result, the parameters have opposite influences on the drag effect. From Eq. (20), $C_1^{dis}$ solely depends on the oscillation frequency of the dislocation core, whereas $C_2^{dis}$ heavily depends on that of the atoms in the perfect system rather than that of the core. This relationship implies that the dislocation core directly causes drag to its motion, which manifests as $C_1^{dis}$, but the remainder of the system aside from the core region tries to resist the drag, which is represented by positive $C_2^{dis}$.

Interestingly, we observed that increasing $C_1^{dis}$ not only leads to decreasing dislocation speed but also increases the nonlinearity of the relationship between speed and applied stress. This is shown in Fig. 12. Given that it has been reported that the nonlinear behavior of defect speed is closely related to the energy dispersion of the moving defect (Eshelby, 1956), our work draws a cogent conclusion.

## 6. Applications

In this section, we solve Eq. (21) and Eq. (26) derived in the preceding section to obtain the speeds of dislocations and LAGBs, respectively. To solve these equations, the two group parameters $C_1^{dis}$ and $C_2^{dis}$ (or $C_1^{LAGB}$ and $C_2^{LAGB}$) should first be calculated. However, it is difficult to calculate all the parameters required to obtain both group parameters since the dislocation core structures are complex and the lattice structure is not as simple as we assumed in modeling. Instead, we avoid this difficulty by directly obtaining both group parameters without calculating their component parameters by fitting the results of molecular dynamics simulations. However, we could roughly estimate a range for $C_2^{dis}/C_1^{dis}$ for oscillating dislocations since the relationship $C_2^{dis}/C_1^{dis} = \left(\pi\Omega/\sqrt{2}w_0\right)^2$ is satisfied. In a 2D system, $0 < \Omega/w_0 < \sqrt{2}$ can be assumed without strong oscillating sources, and thus $C_2^{dis}/C_1^{dis}$ is within $(0, \pi^2)$.

### 6.1. Perfect dislocation in BCC crystals at 0 K

In iron and molybdenum, a dislocation moves as a straight line without dissociation. The speed of dislocations at 0 K as a function of applied stress is shown in Fig 13. As shown in Fig. 13(a) and (b), the slope of the graph decreases as the applied stress increases. This is because the spontaneous oscillation of the dislocation core caused by the relativistic effect dissipates the core energy during its translation (Kim et al., 2019) in addition to the scattering of radiated phonons from the core. As the dislocation velocity increases, the kinetic energy of dislocation increases and the increasing energy is converted into heat through the spontaneous oscillation of the core (Krasnikov and Mayer, 2018). As the applied stress increases, the dislocation speed gradually approaches the transverse shear wave speed, and the oscillation effect becomes more significant. Based on this oscillation mechanism, we solve Eq. (21) by inserting the material properties. First, we calculated the atomic frequency, $w_0$, by $2\sqrt{K/M}$. Here, the spring constant $K$ was calculated from the shear modulus, $G$, through $K = GL_xL_z/L_y$, where $L_x$, $L_y$, and $L_z$ are the dimensions of the simulation cell. We calculated the mobility of dislocation $M_{dis}$ by measuring $dv/d\sigma_{app}$ at $\sigma_{app} \approx \sigma_P$. Then, we inserted proper values of $C_1^{dis}$ and $C_2^{dis}$ into Eq. (21) and obtained the dislocation speed as a function of applied stress. All the parameters are summarized in Table 1. Based on Fig. 13(a), when $C_1^{dis} = 0.910$ and $C_2^{dis} = 1.004$, our theoretical model described the simulation results with a high accuracy for iron. Furthermore, from Fig. 13(b), there is excellent agreement between the model and simulation results when $C_1^{dis} = 0.920$ and $C_2^{dis} = 1.002$ for molybdenum. Interestingly, both group parameters have almost the same values in both materials despite the difference in interatomic potential.

### 6.2. Extended dislocation in FCC crystals at 0 K

In this work, we simulated extended dislocations in aluminum, copper, nickel, and gold. Since the extended dislocation consists of two partial dislocations, they interact with each other, which causes a forced oscillation during gliding motion. Thus, even in the low-speed regime, they are expected to oscillate. As the applied stress increases, the dislocation speed increases, and therefore an additional relativistic effect is included. Thus, two types of oscillations are mixed. By employing the same method discussed in Section 6.1, we calculated and measured the relevant parameters and inserted them with proper $C_1^{dis}$ and $C_2^{dis}$ to solve Eq. (21). The parameters are summarized in Table 1. Then, the solutions were compared with the simulation results in Fig. 14. For aluminum, as shown in Fig. 14(a), when $C_1^{dis} = 0.96$ and $C_2^{dis} = 1.04$, an excellent agreement between the model and simulation results was observed. In addition to aluminum, as shown in Fig. 14(b), (c), and (d), there was remarkable agreement between the simulation results and theory for copper, nickel, and gold. Interestingly, as in BCC crystals, we found that all the extended dislocations in FCC crystals have almost the same $C_1^{dis}$ and $C_2^{dis}$, which are within 0.96-0.97 and 1.02-1.04, respectively.

### 6.3. Dislocations in BCC and FCC crystals at 300 K

In addition to 0 K, we increased the temperature to 300 K and observed the motion of the dislocations at the elevated temperature. As shown in Fig. 15, the speeds of the dislocations at 300 K are much higher than those at 0 K for all crystals under the same stress. The parameters were recalculated and summarized in Table 2. At a finite temperature, a forced oscillation is added that causes both $C_1^{dis}$ and $C_2^{dis}$ to increase, as shown in Table 2 and Fig. 15. As a result, according to Eq. (38) with $w_0 b \ll 2v$, increasing temperature causes the oscillation drag force to increase. During the motion of dislocations at this elevated temperature, we also observed that the cores contracted into more compact structures. This is described in Fig. 16. Therefore, we proved that $J_{dis}$ increases at a high temperature, which causes both group parameters to increase. From Fig. 15, our theoretical model fitted the simulation results with a high accuracy for all cases. Especially, much larger $C_1^{dis}$ and $C_2^{dis}$ are required for nickel at 300 K than for the other crystals. This is because the extended dislocation shrank more rapidly in nickel than in the others.

The contraction of dislocation cores as speed increases has been observed in previous simulation studies (Wang and Beyerlein, 2008; Peng et al., 2019) and theoretically derived to account for relativistic effects and radiation drag during the motion of dislocations (Pellegrini, 2012, 2014). According to Pellegrini (2012), the dependence of core width, $a(v)$, on dislocation speed is given by $a(v) = 2d\left|L(v+i0^+)/2e_0 + i\alpha(v/C_t)\right|$, where $d$ is the interatomic plane separation, $L(v)$ is a stationary Lagrangian, $e_0 (= Gb^2/4\pi)$ is the characteristic energy per unit length of the dislocation line, and $\alpha$ is the drag coefficient. Here, the Lagrangian is derived as $L_s(v) = -e_0 \beta_t$ and $L_e(v) = -4e_0(C_t/v)^2(\beta_l - \beta_t^{-1}\beta_{t_2}^4)$, where $\beta_t = \sqrt{1-(v/C_t)^2}$, $\beta_l = \sqrt{1-(v/C_l)^2}$, and $\beta_{t_2} = \sqrt{1-v^2/(2C_t^2)}$, for screw and edge dislocations, respectively (Hirth et al., 1998). Additionally, as $v \to 0$, $L_s(0) = -e_0$ and $L_e(v) = -e_0/(1-\upsilon)$, where $\upsilon$ is Poisson's ratio, are satisfied. Since both edge and screw dislocations show the same tendency in the core width within the subsonic regime (Pellegrini, 2012), hereafter, we discuss only screw dislocations for an intuitive grasp. Hirth (1998) showed that the total (self) energy of a uniformly moving screw dislocation, $E_s(v)$, satisfies $E_s(v) = e_0 \beta_t^{-1}$. Thus, as the dislocation speed increases, the total energy of the dislocation increases, and the magnitude of the Lagrangian decreases. Finally, this leads to a decrease in core width. Therefore, we conclude that the contraction of the dislocation core at a high temperature observed in this study is a result of increasing dislocation self-energy.

### 6.4. LAGBs at 0 K

Considering that the dislocations comprising LAGBs interact with each other and thereby cause drag to the LAGB motion, the LAGB motion can be regarded as an extension of the motion of extended dislocations introduced in the previous section. However, each dislocation interacts with all other dislocations, not just with a single one, and thus the motion of LAGBs is more complicated than that of extended dislocations. It is also expected that the interactions within the LAGB become a stronger drag source than in the single dislocation case.

The oscillation of a curved LAGB can be regarded as an oscillating string, whose length is $L_{string}$, effective line density is $\rho_{LAGB}^*$, and line tension per thickness is $T_{LAGB}^*$, as shown in Fig. 17(a). The fundamental frequency of the string, $\Omega_{LAGB}$, is expressed as $\Omega_{LAGB} = (1/2L_{string})\sqrt{T_{LAGB}^*/\rho_{LAGB}^*}$. Here, since the effective line density is directly proportional to the effective mass of the LAGB, $m_{LAGB}^*$, we roughly estimate the frequency of the LAGB by deriving its effective mass. Note that this is not a real mass but a quantity that is 'physically' similar to the mass. Effective mass has been defined in various forms depending on its derivation (Brailsford, 1966b; Sakamoto, 1991; Hirth et al., 1998). In this study, we mainly follow the method developed by Brailsford (1966a, 1966b) given that the discreteness of the lattice is considered. However, since his method can be applied only to single dislocations, we extended it to apply to LAGBs with some modifications. We assumed that the LAGB consists of $2N+1$ dislocations embedded in a medium of density $\rho$. As a result, the effective mass of the LAGB is derived as

$$m_{LAGb}^* = (2N^2 + 4N + 1)m_{dis}^* + \frac{\rho b^2}{4\pi^2} \sum_{k=1}^{2N} \sum_{n=1}^{\infty} (2N+1-k)\frac{(-1)^n(q_D kd)^{2n}}{n(2n)!} \int_0^{2\pi} (f(\theta))^{2n} g(\theta)d\theta,$$

(39)

where

$$m_{dis}^* = (\rho b^2/4\pi)(1+r)\ln(q_D/\sigma),$$
(40-1)

$$f(\theta) = \sin\theta + (v/C_t)\cos\theta, \qquad (40\text{-}2)$$

$$g(\theta) = 1 - (1-r)\sin^2 2\theta. \qquad (40\text{-}3)$$

Here, $m_{dis}^*$ is the effective mass of a single edge dislocation, $d$ is the distance between neighboring dislocations, and $q_D$ is the Debye cutoff, which is equal to $(6\pi^2/\Gamma)^{1/3}$, where $\Gamma$ is the atomic volume in real space. A detailed derivation of Eq. (39) is introduced in Appendix C. According to Eq. (39), the effective mass of the LAGB is not just a simple sum of that of its dislocations, or $(2N+1)m_{dis}^*$. Rather, the LAGB has a much larger effective mass that is proportional to the square of the number of dislocations with an extra term. This is because each dislocation interacts not only with its nearest neighbors but also with all other dislocations that comprise the LAGB. This complicates interactions between dislocations, slowing the LAGB and leading to a significant decrease in its oscillation frequency, $\Omega_{LAGB}$.

Since the relationship $C_2^{LAGB}/C_1^{LAGB} = (\pi\Omega_{LAGB}/\sqrt{2}w_0)^2$ is still satisfied for the LAGB, we expect that $C_2^{LAGB}/C_1^{LAGB}$ is much smaller than for a single dislocation. All the parameters to solve Eq. (26) are summarized in Table 3. Here, $M_{dis}$ and $w_0$ were computed from an atomistic simulation of a single dislocation and perfect lattice using the same sized simulation box and LJ potential parameters, respectively. These values were used in our previous work (Kim et al., 2019). $C_1^{LAGB}$ and $C_2^{LAGB}$ were used as phenomenological parameters that depend on applied stress to compare with the simulation results. Strictly speaking, $C_2^{LAGB}/C_1^{LAGB}$ depends on both misorientation angle and applied stress, but we assume that the ratio changes mainly as a result of changes in applied stress because the difference between the minimum and maximum misorientation angles used in this work is less than $5°$, which is an extremely small value. Thus, we expect that this simplification does not influence the main result of our work. According to Wei and Peng (2017), they used the frequency of transverse phonons—which is the main source of radiation drag during dislocation motion in the subsonic regime—as a phenomenological parameter to fit dislocation travel distance over time. In their work, they used a larger frequency as the applied strain rate increases. With reference to their results, we assumed that $\Omega_{LAGB}$ increases with increasing applied stress in this study, which causes $C_1^{LAGB}$ to decrease and $C_2^{LAGB}$ to increase according to Eq. (24). This is reflected in Table 3. As a result, we inversely calculated $\Omega_{LAGB}$ by the relationship $\Omega_{LAGB} = (w_0/\pi)\sqrt{2C_2^{LAGB}/C_1^{LAGB}}$ using the values listed in Table 3 to be within the range of $1.51 \sim 1.60$ THz. Given that this range of frequencies is similar to the frequency spectrum in which a dislocation array acts as an effective phonon scattering source for lattice heat conduction (Kim et al., 2015; Kim et al., 2016), our estimate of $C_1^{LAGB}$ and $C_2^{LAGB}$ is reasonable.

A comparison between the solution of Eq. (26) and simulation results is shown in Fig. 18. As a result, for the valid $C_1^{LAGB}$ and $C_2^{LAGB}$, our theoretical model describes an inverse relationship between the speed and misorientation angle of LAGBs, which cannot be explained in terms of linear elasticity theory, with high accuracy. Therefore, it proves that the phonon drag effect occurring due to elastic interactions among dislocations comprising the LAGB is responsible for this relationship. This effect should be considered when the LAGB moves in a dispersive medium.

## 7. Conclusions

While a defect moves in a discrete medium, unusual behaviors that are not fully explainable by linear elasticity theory were observed through atomistic simulations. We proved that a dynamic drag effect is responsible for these behaviors. During dislocation motion, a stress-drop phenomenon was observed owing to phonon scattering around the core. As the speed of the dislocation gradually increases, or as it interacts with other dislocations, the dislocation core oscillates. These oscillations induce additional phonon drag to the motion of the dislocation. In addition to stress-drop, two unusual phenomena were observed during the motion of LAGBs. First, the speed was inversely proportional to the misorientation angle. Second, the LAGB maintained a curved form, even in an equilibrium state. Both results cannot be explained solely by the linear elasticity theory. However, by introducing the drag effect originating from interactions among constituent dislocations of the LAGB, both unusual behaviors could be explained.

The phonon drag effect not only explains the unusual behaviors of defects but also generally governs the defect dynamics in a discrete system. In this work, we developed a general equation of motion for dislocations where the drag effect is considered. The equation was derived based on the DLD theory by assuming that dislocations oscillate by various sources. There was excellent agreement between our theoretical model and the results of atomistic simulations when proper group parameters were used. Interestingly, even if the materials were different, the group parameters were almost the same as long as the lattice system did not change. Given that the group parameters quantify the magnitude of the drag effect, this proves that the drag effect depends on the *geometric structure* of the dislocation core rather than the material properties. Furthermore, the simulations showed that a contraction of the dislocation cores occurs in cubic crystals as temperature increases, leading to an increase in the drag effect via an increase in the group parameters. Despite the good agreement between the theoretical model and simulations, our model has a weakness in that the group parameters are still determined via a phenomenological process. By directly deriving the parameters in our future work, we will extend the model to cases in which the drag is more severely generated by obstacles such as Frank–Read sources, triple junctions, etc.

In addition to dislocations, we could extend our theory to LAGBs given that they oscillate in a manner similar to dislocations. Especially, since LAGBs consist of a number of dislocations with more complicated interactions than that of an extended dislocation, the drag effect on the LAGB motion is characterized by a significant decrease in its oscillation frequency. Based on this analysis, we solved the equation by inserting the roughly estimated group parameters and fitted the solutions to the simulation results. As a result, the inverse relationship between the speed and misorientation angle of LAGBs could be explained with high accuracy.

Our theoretical model has academic significance in that it can be *generally* applied to moving defects despite the diversity of the drag sources. This is possible because the drag effect manifests as an oscillation of the cores as a common result. This can be simply quantified in terms of the two grouping parameters that are defined in this study with a thorough analysis. Therefore, we believe that this study will contribute to understanding the fundamental dynamics of defects in discrete dispersive media.

**Acknowledgement**

We gratefully acknowledge the support from the Mid-Career Researcher Support Program (Grant No. 2019R1A2C2011312) of the National Research Foundation (NRF) of Korea and from the High-Speed Manufacturing and Commercialization of Ultralight Composites Research Fund (Project No. 1.190009.01) of UNIST. We also acknowledge with gratitude the supercomputing resources of the UNIST Supercomputing Center.

**Appendix A. Minimum energy state of the LAGB based on linear elasticity theory**

Assume that $N$ dislocations that they were initially at $(x_0, y_i)$, where $i = 1, 2, \cdots, N$, move positive $x$ direction without cross-slip. At arbitrary time, $t$, the total energy of the system, $E_{tot}$, can be expressed by summation of elastic strain energy done by the externally applied stress, $E_{app}$, self-energy of each dislocation, $E_{self}$, and interaction energy between the dislocations, $E_{int}$, as Eq. (A.1).

$$E_{tot} = E_{app} + NE_{self} + E_{int}. \tag{A.1}$$

At dynamic equilibrium state, $E_{tot}$ must be minimized with respect to dislocation position at $t$, $x_i(t)$. Since $E_{app}$ and $E_{self}$ are independent to the dislocation position, $E_{tot}$ has minimum value when $E_{int}$ is minimized.

The shear stress acting at $(x, y)$ due to $i$ th dislocation at $(x_i, y_i)$, whose Burgers vector is $\mathbf{b}$, is expressed as Eq. (A.2).

$$\sigma_{12}(x, y) = \frac{Gb}{2\pi(1-v)} \frac{(x-x_i)[(x-x_i)^2 - (y-y_i)^2]}{[(x-x_i)^2 + (y-y_i)^2]^2} \tag{A.2}$$

where $G$ is a shear modulus, $v$ is Poisson's ratio. Thus, the shear stress acting at $j$ th dislocation induced by interaction with other $N-1$ dislocations becomes

$$\sigma_{12}^{j} = \frac{Gb}{2\pi(1-v)} \sum_{\substack{i=1 \\ (i \neq j)}}^{N} \frac{(x_j - x_i)[(x_j - x_i)^2 - (y_j - y_i)^2]}{[(x_j - x_i)^2 + (y_j - y_i)^2]^2}. \tag{A.3}$$

The interaction energy between the $j$ th dislocation and other dislocations, $E_{int}^{j}$, is a work done by displacing the cut plane, $y = y_j$, to make the $j$ th dislocation in the presence of stress field $\sigma_{12}^{j}$ is derived as Eq. (A.4).

$$E_{int}^{j} = \int_{x_j}^{\infty} \sigma_{12}^{j} b dx_j. \tag{A.4}$$

Therefore, the total interaction energy can be derived as

$$E_{int} = \frac{1}{2} \sum_{j=1}^{N} E_{int}^{j} = \frac{Gb}{4\pi(1-v)} \sum_{j=1}^{N} \sum_{\substack{i=1 \\ (i \neq j)}}^{N} \int_{x_j}^{\infty} \frac{(x_j - x_i)[(x_j - x_i)^2 - (y_j - y_i)^2]}{[(x_j - x_i)^2 - (y_j - y_i)^2]^2} dx_j. \tag{A.5}$$

To find the equilibrium state, both $\partial E_{int} / \partial x_j = 0$ and $\partial^2 E_{int} / \partial x_j^2 > 0$ must be satisfied. As a result, when $x_i = x_j (i = 1, 2, \cdots, N, (i \neq j))$, the both conditions are satisfied. In other words, when the LAGB maintains its configuration as a straight line, it has a minimum energy.

**Appendix B. Derivation of lattice Green's function**

The lattice Green's function in one-dimensional lattice with considering an anharmonic atomic displacement was firstly derived by Ohashi (1968). We expanded it to two-dimensional case in our previous study (Kim et al., 2016) and reproduce it in here. If we define a displacement of atom $m$ in response to elastic stress, $V_m$, and additional deviation from the static equilibrium position due to anharmonic scattering as $u_m$, then total displacement becomes $V_m + u_m$. The total energy of the system, $H$, can be expressed by

$$H = \frac{1}{2} M \sum_n \dot{u}_n^2 + \sum_n A_n (V_n + u_n) + \frac{1}{2} \sum_m \sum_n B_{mn}(V_m + u_m)(V_n + u_n)$$
$$+ \frac{1}{6} \sum_l \sum_m \sum_n C_{lmn}(V_l + u_l)(V_m + u_m)(V_n + u_n) + O(V^4, u^4)$$
(B.1)

where $A_n, B_{mn}$, and $C_{lmn}$ are first, second, and third derivatives of potential energy in unconstrained state, $E$, respectively. In here, $A_n$ is equal to zero because the potential energy becomes minimum in equilibrium state and $B_{mn}$ corresponds to an elastic constant. If we consider a potential energy occurred by only elastic displacement, $E_o$, it is expressed as

$$E_0 = \frac{1}{2} \sum_m \sum_n B_{mn} V_m V_n + \frac{1}{6} \sum_l \sum_m \sum_n C_{lmn} V_l V_m V_n + O(V^4).$$
(B.2)

Since harmonic displacement minimizes system's potential energy, $\partial E_0 / \partial V_n$ should be zero. Thus,

$$\sum_m B_{mn} V_m + \frac{1}{2} \sum_l \sum_m C_{lm} V_l V_m + O(V^3) = 0.$$
(B.3)

If we insert Eq. (B.3) into Eq. (B.1) and ignore the high-order terms, the total potential energy of the system becomes

$$E = E_0 + \frac{1}{2} \sum_m \sum_n B_{mn} u_m u_n + \frac{1}{2} \sum_l \sum_m \sum_n C_{lmn} V_l u_m u_n.$$
(B.4)

Therefore, the contribution of anharmonicity to system's total energy becomes

$$H_{anh} = \frac{1}{2} M \sum_m \dot{u}_m^2 + \frac{1}{2} \sum_m \sum_n B_{mn} u_m u_n + \frac{1}{2} \sum_l \sum_m \sum_n C_{lmn} V_l u_m u_n .\tag{B.5}$$

From $\partial H_{anh} / \partial t = 0$, the equation of motion is derived as Eq. (B.6).

$$M \ddot{u}_m \delta_{mn} + \sum_m B_{mn} u_m + \sum_l \sum_m C_{lmn} V_l u_m = 0 .\tag{B.6}$$

If we insert $u_m = u_0 \exp(-iwt)$, then Eq. (B6) is converted to Eq. (B.7).

$$\sum_m [-Mw^2 \delta_{mn} + B_{mn}] u_m = -\sum_m \left[ \sum_l C_{lmn} V_l \right] u_m .\tag{B.7}$$

In matrix equation, Eq. (B.7) becomes

$$\mathbf{L}\vec{u} = \Delta \mathbf{L}\vec{u} \tag{B.8}$$

where $L_{mn} = -Mw^2 \delta_{mn} + B_{mn} = M(w^2(k) - w^2)\delta_{mn}$ and $\Delta L_{mn} = -\sum_l C_{lmn} V_l$. Here, $w(k)$ is a dispersion of waves. The solution of Eq. (B.8) can be expressed as a sum of incoming and scattered waves, which are defined as $\vec{u}_i$ and $\vec{u}_s$, respectively. Since $\vec{u}_i$ is a homogenous but $\vec{u}_s$ is a particular solution, the solution of Eq. (B.8) can be derived as Eq. (B.9).

$$\vec{u} = \vec{u}_i + \mathbf{L}^{-1} \Delta \mathbf{L} \vec{u} .\tag{B.9}$$

If we define the lattice Green's function, $\mathbf{G}$, as $\mathbf{G} \equiv \mathbf{L}^{-1}$. Thus, $\mathbf{G}$ is derived as Eq. (B.10) in first Brillouine zone (BZ).

$$G_{mn}(w) = \frac{b}{2\pi M} \int_{1st\,BZ} \frac{e^{ikb(m-n)}}{w^2(k) - w^2} dk .\tag{B.10}$$

In this work, however, we apply $\mathbf{G}$ in two-dimensional lattice so it is defined as Eq. (B.11).

$$G(\vec{r} - \vec{r}_0, w) = \frac{b^2}{4\pi^2 M} \int_{1st\, BZ} \frac{e^{i\vec{k}\cdot(\vec{r}-\vec{r}_0)}}{w^2(\vec{k}) - w^2} d\vec{k}. \tag{B.11}$$

**Appendix C. Derivation of effective mass of the LAGB**

The displacement at $\mathbf{r}$, $\mathbf{u}(\mathbf{r})$, can be described in complex plane as

$$\mathbf{u}(\mathbf{r}) = \sum_{\mathbf{q}} \sum_{p} u_{\mathbf{q}p} e^{i\mathbf{q}\cdot\mathbf{r}} \hat{\mathbf{e}}_{\mathbf{q}p} \tag{C.1}$$

where $\hat{\mathbf{e}}_{\mathbf{q}p}$ is a polarization vector for three modes ($p = 1, 2, 3$), and $q$ is a wavenumber. Here, $p = 1$ corresponds to longitudinal and $p = 2$ and $3$ correspond to transverse modes of the dislocation. The dislocation is embedded as in Fig. 17 and its slip area is confined to $z = 0$ plane. Additionally, $\bar{\mathbf{u}}_{\mathbf{q}p} = \mathbf{u}_{-\mathbf{q}p}$ is satisfied.

The kinetic energy of system is derived as

$$T = \frac{1}{2}\rho V \left|\frac{d\mathbf{u}}{dt}\right|^2 = \frac{1}{2}\rho V \sum_{\mathbf{q}} \sum_{p} \bar{\mathbf{u}}_{\mathbf{q}p} \mathbf{u}_{\mathbf{q}p} \tag{C.2}$$

where $\rho$ and $V$ is density and volume of the system, respectively. Elastic strain energy is expressed as

$$U_{elastic} = \frac{1}{2}V \sum_{p} C_p \left|\frac{\partial \mathbf{u}}{\partial \mathbf{r}}\right|^2 = \frac{1}{2}V \sum_{\mathbf{q}} \sum_{p} q^2 C_p \bar{\mathbf{u}}_{\mathbf{q}p} \mathbf{u}_{\mathbf{q}p} \tag{C.3}$$

where $C_1 = 2G(1-\upsilon)/(1-2\upsilon)$, and $C_2 = C_3 = G$. Here, $\upsilon$ is Poisson's ratio.

And the potential energy of LAGB is derived as

$$U_{LAGB} = \frac{1}{2}\sum_{\mathbf{R}}\int_C \mathbf{n}\cdot\boldsymbol{\sigma}(\mathbf{r}-\mathbf{R})\cdot\mathbf{b}\,dS = \frac{1}{2}\int_C n_i\left(\sum_{\mathbf{R}}\sigma_{ij}(\mathbf{r}-\mathbf{R})\right)b_j\,dS, \tag{C.4}$$

where $S$ is slipped area by each dislocation whose Burgers vector is $\mathbf{b}$ and is located at $\mathbf{R}$, and $\mathbf{n}$ is normal vector the area. Since only shear components of the stress contributes to $U_{LAGB}$, the stress is derived as

$$\boldsymbol{\sigma}(\mathbf{r}) = G\frac{\partial \mathbf{u}}{\partial \mathbf{r}} = iG\sum_{\mathbf{q}}\sum_{p}\left(\hat{\mathbf{q}}\hat{\mathbf{e}}_{\mathbf{q}p} + \hat{\mathbf{e}}_{\mathbf{q}p}\hat{\mathbf{q}}\right)qu_{\mathbf{q}p}e^{i\mathbf{q}\cdot\mathbf{r}}, \tag{C.5}$$

where $\hat{\mathbf{q}}$ is a normalized vector of $\mathbf{q}$. If we insert Eq. (C.5) into Eq. (C.4),

$$U_{LAGB} = \frac{1}{2}\sum_{\mathbf{q}}\sum_{p} A_{\mathbf{q}p}\left(\sum_{\mathbf{R}} e^{-i\mathbf{q}\cdot\mathbf{R}}\right)u_{\mathbf{q}p}, \tag{C.6}$$

where

$$A_{\mathbf{q}p} = Gbq\Phi_{\mathbf{q}p}f(\mathbf{q}), \tag{C.7}$$

$$\Phi_{\mathbf{q}p} = (\hat{\mathbf{n}}\cdot\hat{\mathbf{q}})(\hat{\mathbf{e}}_{\mathbf{q}p}\cdot\hat{\mathbf{b}}) + (\hat{\mathbf{n}}\cdot\hat{\mathbf{e}}_{\mathbf{q}p})(\hat{\mathbf{q}}\cdot\hat{\mathbf{b}}), \tag{C.8}$$

$$f(\mathbf{q}) = i\oint_C e^{i\mathbf{q}\cdot\mathbf{r}}\,dS. \tag{C.9}$$

Here, $\hat{\mathbf{b}}$ is a normalized vector of $\mathbf{b}$. Furthermore, since elastic wave propagates from the free edges of system, each dislocation that consists the LAGB starts to move at different time. As a result, if we assume that $0^{th}$ dislocation starts to move at $t = 0$, $j$-th dislocation starts its motion at $t = -\frac{jd}{C_t}$ where $d$ is a distance between neighboring dislocations and $C_t$ is a speed of the shear wave. Therefore, $A_{\mathbf{q}p}$ becomes

$$A_{\mathbf{q}p}(\mathbf{q},\mathbf{R},\mathbf{v},t) = Gbq\Phi_{\mathbf{q}p}\left(i\int_C e^{-i\mathbf{q}\cdot(\mathbf{r}-\mathbf{v}t)}\,dS\right) = A_{\mathbf{q}p}^0 e^{-i\mathbf{q}\cdot\mathbf{v}t} e^{-i\mathbf{q}\cdot\mathbf{v}\frac{R}{C_t}}, \qquad (C.10)$$

where $R = jd$. This leads $U_{LAGB}$ to become

$$U_{LAGB} = \frac{1}{2}\sum_{\mathbf{q}}\sum_p A_{\mathbf{q}p}^0 \left[\sum_R e^{-i\mathbf{q}\cdot R\left(\hat{\mathbf{e}}_3 + \frac{\mathbf{v}}{C_t}\right)}\right] e^{-i\mathbf{q}\cdot\mathbf{v}t} u_{\mathbf{q}p} = \frac{1}{2}\sum_{\mathbf{q}}\sum_p B_{\mathbf{q}p}(\mathbf{q},\mathbf{v},t) u_{\mathbf{q}p}, \qquad (C.11)$$

where $B_{\mathbf{q}p} = A_{\mathbf{q}p}^0 \sum_R \left[e^{-i\mathbf{q}\cdot R(\hat{\mathbf{e}}_3 + \mathbf{v}/C_t)}\right] e^{-i\mathbf{q}\cdot\mathbf{v}t}$.

Therefore, Lagrangian, $L$, becomes $L = T - (U_{elastic} + U_{LAGB})$. Through Euler-Lagrange equation in complex space, it is reduced to

$$\ddot{u}_{\mathbf{q}p} + w_{\mathbf{q}p}^2 u_{\mathbf{q}p} + \frac{1}{\rho V}\bar{B}_{\mathbf{q}p} = 0, \qquad (C.12)$$

where $w_{\mathbf{q}p} = q\sqrt{C_p/\rho}$. The solution of Eq. (C.12) is expressed by a sum of homogenous and particular solutions as

$$u_{\mathbf{q}p} = u_{\mathbf{q}p}^h(t) + u_{\mathbf{q}p}^p(t) = c_{\mathbf{q}p}^1 e^{iw_{\mathbf{q}p}t} + c_{\mathbf{q}p}^2 e^{-iw_{\mathbf{q}p}t} - \frac{\bar{B}_{\mathbf{q}p}(t)}{\rho V\{w_{\mathbf{q}p}^2 - (\mathbf{q}\cdot\mathbf{v})^2\}}. \qquad (C.13)$$

In addition to the energy of free oscillators, the particular solution induces kinetic energy occurred by uniform motion of the LAGB, $T_{LAGB}$. It becomes

$$T_{LAGB} = \frac{1}{2}\rho V \sum_{\mathbf{q}}\sum_{p} \bar{\dot{u}}_{\mathbf{q}p}^{p} \dot{u}_{\mathbf{q}p}^{p}$$

$$= \frac{1}{2}\frac{1}{\rho V}\sum_{\mathbf{q}}\sum_{p}(\mathbf{q}\cdot\mathbf{v})^2 \frac{\bar{B}_{\mathbf{q}p}B_{\mathbf{q}p}}{\{w_{\mathbf{q}p}^2-(\mathbf{q}\cdot\mathbf{v})^2\}^2}$$

$$= \frac{1}{2}\frac{1}{\rho V}\sum_{\mathbf{q}}\sum_{p}(\mathbf{q}\cdot\mathbf{v})^2 \frac{|B_{\mathbf{q}p}|^2}{w_{\mathbf{q}p}^4}\left\{1-\left(\frac{\mathbf{q}\cdot\mathbf{v}}{w_{\mathbf{q}p}}\right)^2\right\}^{-2} \quad (C.14)$$

$$\approx \frac{1}{2}\frac{1}{\rho V}\sum_{\mathbf{q}}\sum_{p}(\mathbf{q}\cdot\mathbf{v})^2 \frac{|B_{\mathbf{q}p}|^2}{w_{\mathbf{q}p}^4}\left\{1+2\left(\frac{\mathbf{q}\cdot\mathbf{v}}{w_{\mathbf{q}p}}\right)^2\right\}.$$

Also, additional potential energy becomes $U_{add} \sim \sum_{\mathbf{q}}\sum_{p} \bar{u}_{\mathbf{q}p}^{p} u_{\mathbf{q}p}^{p}$ so that

$$U_{add} \sim \frac{1}{(\rho V)^2}\sum_{\mathbf{q}}\sum_{p}\frac{|B_{\mathbf{q}p}|^2}{w_{\mathbf{q}p}^4}\left\{1+2\left(\frac{\mathbf{q}\cdot\mathbf{v}}{w_{\mathbf{q}p}}\right)^2\right\}. \quad (C.15)$$

By comparing the lowest-order term of $\mathbf{v}$ in between Eq. (C.14) and Eq. (C.15), it is concluded that the latter is negligible compared to the former. Therefore, it is valid to consider only kinetic term. As a result,

$$T_{LAGB} = \frac{1}{2}\frac{1}{\rho V}\sum_{\mathbf{q}}\sum_{p}(\mathbf{q}\cdot\mathbf{v})^2 \frac{|B_{\mathbf{q}p}|^2}{w_{\mathbf{q}p}^4} = \frac{1}{2}m_{ij}^{LAGB}v_i v_j \quad (C.16)$$

where $m_{ij}^{LAGB}$ is defined as an effective mass of LAGB. As a result, it is expressed as

$$m_{ij}^{LAGB} = \frac{\rho b^2}{V}\sum_{\mathbf{q}}\sum_{p}\left\{\frac{G\Phi_{\mathbf{q}p}|f(\mathbf{q})|}{C_p}\right\}^2\left[\sum_{R,R'}e^{-i\mathbf{q}\cdot(R-R')\left(\hat{\mathbf{e}}_3+\frac{\mathbf{v}}{C_t}\right)}\right]\hat{q}_i\hat{q}_j. \quad (C.17)$$

This is a general definition of the effective mass. Now, we derive $m_{ij}^{LAGB}$ when the LAGB consists of infinitely-long straight dislocations.

From Fig. 17, we derive $f(\mathbf{q})$ for straight dislocation whose length is $L$ at first. So, Eq. (C.9) becomes Eq. (C.18).

$$f(\mathbf{q}) = i \int_{-\infty}^{0} e^{iq_y y} dy \int_{-L/2}^{L/2} e^{iq_x x} dx$$
$$= \frac{2\sin(q_x L/2)}{q_x} i \int_{-\infty}^{0} e^{iq_y y} dy = \frac{2\sin(q_x L/2)}{q_x} i \int_{0}^{\infty} e^{-iq_y y} dy. \tag{C.18}$$

By Heitler (1944), Eq. (C.18) is converted to

$$f(\mathbf{q}) = \lim_{\sigma \to 0} \frac{2\sin(q_x L/2)}{q_x (q_y - i\sigma)}. \tag{C.19}$$

Since the dislocation moves $y$ direction on $z=0$ plane, the effective mass for $L \to \infty$ becomes

$$m_{yy}^{LAGB} = \frac{\rho b^2}{V} \lim_{\substack{\sigma \to 0 \\ L \to \infty}} \sum_{\mathbf{q}} \sum_{p} (\hat{q}_y)^2 \left\{ \frac{G\Phi_{\mathbf{q}p}}{C_p} \right\}^2 \left| \frac{2\sin(q_x L/2)}{q_x (q_y - i\sigma)} \right|^2 \left[ \sum_{R,R'} e^{-i\mathbf{q}\cdot(R-R')\left(\hat{\mathbf{e}}_3 + \frac{\mathbf{v}}{C_t}\right)} \right]. \tag{C.20}$$

And by using Dirac delta function, $\delta(x)$, Eq. (C.21) is obtained.

$$\lim_{L \to \infty} \left| \frac{2\sin(q_x L/2)}{q_x (q_y - i\sigma)} \right|^2 = \frac{2}{q_y^2 + \sigma^2} \lim_{L \to \infty} \frac{1-\cos(q_x L)}{q_x^2}$$
$$= \frac{2}{\pi} \frac{1}{q_y^2 + \sigma^2} \lim_{L \to \infty} \frac{1}{L} \frac{1-\cos(q_x L)}{q_x^2} (\pi L) \tag{C.21}$$
$$= \frac{2\pi L \delta(q_x)}{q_y^2 + \sigma^2}.$$

If we insert Eq. (C.21) into Eq. (C.20) and replace $\sum_{\mathbf{q}}$ with $\frac{V}{(2\pi)^3} \int d^3\mathbf{q}$, then the effective mass per unit length, $m^*_{LAGB}$, is derived as Eq. (C.22).

$$m_{LAGB}^{*} = \frac{\rho b^{2}}{4\pi^{2}} \sum_{R,R'} \int \left[ (\hat{q}_{y})^{2} \delta(q_{x}) \frac{1}{q_{y}^{2}+\sigma^{2}} \sum_{p} \left\{ \frac{G\Phi_{qp}}{C_{p}} \right\}^{2} \right] \left[ e^{-i\mathbf{q}\cdot(R-R')\left(\hat{\mathbf{e}}_{3}+\frac{\mathbf{v}}{C_{t}}\right)} \right] d\mathbf{q}. \quad \text{(C.22)}$$

By Ehelby (1962), if we define $\boldsymbol{\varphi}_{\mathbf{q}} = (\hat{\mathbf{q}}\cdot\hat{\mathbf{b}})\mathbf{n} + (\hat{\mathbf{q}}\cdot\hat{\mathbf{n}})\hat{\mathbf{b}}$, we can express $\Phi_{qp}^{2}$ from Eq. (C.8).

$$\Phi_{qp}^{2} = \boldsymbol{\varphi}_{\mathbf{q}} \cdot \hat{\mathbf{e}}_{qp}\hat{\mathbf{e}}_{qp} \cdot \boldsymbol{\varphi}_{\mathbf{q}} \quad \text{(C.23)}$$

Therefore,

$$\begin{aligned}
\sum_{p} \left( \frac{G\Phi_{qp}}{C_{p}} \right)^{2} &= r\Phi_{q1}^{2} + \Phi_{q2}^{2} + \Phi_{q3}^{2} \\
&= r\boldsymbol{\varphi}_{\mathbf{q}} \cdot \hat{\mathbf{e}}_{q1}\hat{\mathbf{e}}_{q1} \cdot \boldsymbol{\varphi}_{\mathbf{q}} + \boldsymbol{\varphi}_{\mathbf{q}} \cdot (\tilde{\mathbf{I}} - \hat{\mathbf{e}}_{q1}\hat{\mathbf{e}}_{q1}) \cdot \boldsymbol{\varphi}_{\mathbf{q}} \\
&= \boldsymbol{\varphi}_{\mathbf{q}} \cdot (\tilde{\mathbf{I}} - (1-r)\hat{\mathbf{e}}_{q1}\hat{\mathbf{e}}_{q1}) \cdot \boldsymbol{\varphi}_{\mathbf{q}} \\
&= \varphi_{\mathbf{q}}^{2} - (1-r)(\boldsymbol{\varphi}_{\mathbf{q}} \cdot \hat{\mathbf{e}}_{q1})^{2}.
\end{aligned} \quad \text{(C.24)}$$

where $r = G/C_{1}$ and $\tilde{\mathbf{I}} = \hat{\mathbf{e}}_{q1}\hat{\mathbf{e}}_{q1} + \hat{\mathbf{e}}_{q2}\hat{\mathbf{e}}_{q2} + \hat{\mathbf{e}}_{q3}\hat{\mathbf{e}}_{q3}$ is the identity tensor. Since we define $\hat{\mathbf{e}}_{q1} = \hat{\mathbf{q}}$, Eq. (C.24) is converted to Eq. (C.25).

$$\sum_{p} \left( \frac{G\Phi_{qp}}{C_{p}} \right)^{2} = (\hat{\mathbf{q}}\cdot\hat{\mathbf{b}})^{2} + (\hat{\mathbf{q}}\cdot\hat{\mathbf{n}})^{2} - 4(1-r)(\hat{\mathbf{q}}\cdot\hat{\mathbf{b}})^{2}(\hat{\mathbf{q}}\cdot\hat{\mathbf{n}})^{2}. \quad \text{(C.25)}$$

If we define $\hat{\mathbf{q}} = \hat{q}_{x}\hat{\mathbf{e}}_{1} + \hat{q}_{y}\hat{\mathbf{e}}_{2} + \hat{q}_{z}\hat{\mathbf{e}}_{3}$, $\hat{\mathbf{b}} = \cos\Theta\hat{\mathbf{e}}_{1} + \sin\Theta\hat{\mathbf{e}}_{2}$ and $\hat{\mathbf{n}} = \hat{\mathbf{e}}_{3}$ according to Fig. 17, Eq. (C.25) becomes

$$\sum_{p} \left( \frac{G\Phi_{qp}}{C_{p}} \right)^{2} = (\hat{q}_{x}\cos\Theta + \hat{q}_{y}\sin\Theta)^{2} + \hat{q}_{z}^{2} - 4(1-r)\hat{q}_{z}^{2}(\hat{q}_{x}\cos\Theta + \hat{q}_{y}\sin\Theta)^{2}, \quad \text{(C.26)}$$

where $\Theta$ is an angle between dislocation line and Burgers vector. After inserting Eq. (C.26) to Eq. (C.22), the integration is carried out over a sphere, whose radius is $q_{D}$ that is defined as Debye cutoff. Owing to the existence of $\delta(q_x)$ in the integrand, the integration regime is confined to a circle on $q_x = 0$ plane. So if we define $q_y = q\cos\theta$ and $q_z = q\sin\theta$ where $q = \sqrt{q_y^2 + q_z^2}$ and let $\sigma \to 0$, Eq. (C.22) becomes

$$m^*_{LAGb} = \frac{\rho b^2}{4\pi^2} \int_\sigma^{q_D} \frac{1}{q} dq$$
$$\int_0^{2\pi} \left[\cos^2\theta \sin^2\Theta + \sin^2\theta - 4(1-r)\sin^2\theta \cos^2\theta \sin^2\Theta\right] \left[\sum_{R,R'} e^{-iq(R-R')(\sin\theta + \frac{v}{C_t}\cos\theta)}\right] d\theta.$$

(C.27)

If we define $f(\theta) \equiv \sin\theta + (v/C_t)\cos\theta$, following relation is satisfied.

$$\sum_{R,R'} e^{-iq(R-R')f(\theta)} = \sum_{j=-N}^{N} e^{-iqdjf(\theta)} \sum_{j'=-N}^{N} e^{iqdj'f(\theta)} = \left(\sum_{j'=-N}^{N} e^{iqdj'f(\theta)}\right)^2$$
$$= 2\sum_{k=1}^{2N}(2N+1-k)\cos(qkdf(\theta)) + 2N+1,$$

(C.28)

where $2N+1$ is the total number of dislocations that make up the LAGB. As a result, the effective mass becomes

$$m^*_{LAGB} = \frac{\rho b^2}{4\pi^2} \int_\sigma^{q_D}\int_0^{2\pi} \frac{1}{q}\left[\cos^2\theta \sin^2\Theta + \sin^2\theta - 4(1-r)\sin^2\theta\cos^2\theta\sin^2\Theta\right]$$
$$\left[2N+1+2\sum_{k=1}^{N}(2N+1-k)\cos(gkdf(\theta))\right] dq d\theta.$$

(C.29)

Since all the LAGBs that are considered in our work consist of only edge-type dislocations whose Burgers vector lies on $xy$ plane, $\Theta = \pi/2$ is inserted. Then, Eq. (C.29) is simplified to

$$m^*_{LAGb} = \frac{\rho b^2}{4\pi}(2N+1)(1+r)\ln\left(\frac{q_D}{\sigma}\right)$$
$$+ \frac{\rho b^2}{2\pi^2}\sum_{k=1}^{2N}(2N+1-k)\int_0^{2\pi}\left[1-4(1-r)\sin^2\theta\cos^2\theta\right]d\theta\int_0^{q_D}\frac{1}{q}\cos(qkdf(\theta))dq. \quad \text{(C.30)}$$

If we use following relation

$$\int\frac{\cos(x)}{x}dx = \frac{1}{2}\sum_{n=1}^{\infty}\frac{(-1)^n x^{2n}}{n(2n)!} + \ln(x) + \gamma + const, \quad \text{(C.31)}$$

where $\gamma$ is Euler-Mascheroni constant, the Eq. (C.30) becomes

$$m^*_{LAGB} = \frac{\rho b^2}{4\pi}(2N+1)(1+r)\ln\left(\frac{q_D}{\sigma}\right) + \frac{\rho b^2}{2\pi}(1+r)\sum_{k=1}^{2N}(2N+1-k)\ln\left(\frac{q_D}{\sigma}\right)$$
$$+ \frac{\rho b^2}{4\pi^2}\sum_{k=1}^{2N}\sum_{n=1}^{\infty}(2N+1-k)\frac{(-1)^n(q_D kd)^{2n}}{n(2n)!}\int_0^{2\pi}(f(\theta))^{2n}\left[1-4(1-r)\sin^2\theta\cos^2\theta\right]d\theta.$$

(C.32)

Since $\sum_{k=1}^{2N}(2N+1-k) = N(N+1)$, the effective mass of LAGB is finally derived as

$$m^*_{LAGb} = \frac{\rho b^2}{4\pi}(1+r)(2N^2+4N+1)\ln\left(\frac{q_D}{\sigma}\right)$$
$$+ \frac{\rho b^2}{4\pi^2}\sum_{k=1}^{2N}\sum_{n=1}^{\infty}(2N+1-k)\frac{(-1)^n(q_D kd)^{2n}}{n(2n)!}\int_0^{2\pi}(f(\theta))^{2n}\left[1-(1-r)\sin^2 2\theta\right]d\theta$$
$$= (2N^2+4N+1)m^*_{dis} + \frac{\rho b^2}{4\pi^2}\sum_{k=1}^{2N}\sum_{n=1}^{\infty}(2N+1-k)\frac{(-1)^n(q_D kd)^{2n}}{n(2n)!}\int_0^{2\pi}(f(\theta))^{2n}g(\theta)d\theta,$$

(C.33)

where $m^*_{dis} = (\rho b^2/4\pi)(1+r)\ln(q_D/\sigma)$ and $g(\theta) = 1-(1-r)\sin^2 2\theta$. Here $m^*_{dis}$ is effective mass of single edge dislocation, which was derived by Brailsford (1965b). And since $q_D$ is defined in discrete momentum space, the system has finite number of degree of freedoms.

Therefore, $q_D$ is defined as $\left(6\pi^2/\Gamma\right)^{1/3}$ where $\Gamma$ is an atomic volume in real space.

## References

Ackland, G. J. Tichy, G., Vitek, V., Finnis, M. W., 2006. Simple N-body potentials for the noble metals and nickel. Philosophical Magazine A 56, 735-756. https://doi.org/10.1080/01418618708204485

Amrit, J., Ramiere, A., Volz, S., 2018. Role of fluttering dislocations in the thermal interface resistance between silicon crystal and plastic solid $^4$He. Physical Review B 97, 014308. https://doi.org/10.1103/PhysRevB.97.014308

Angelo, J. E., Moody, N. R., Baskes, M. I., 1995. Trapping of hydrogen to lattice defects in nickel. Modelling and Simulation in Materials Science and Engineering 3, 289-307. https://doi.org/10.1088/0965-0393/3/3/001

Atkinson, W., Cabrera, N., 1965. Motion of a Frenkel-Kontorowa dislocation in a one-dimensional crystal. Physical Review 138, 763-766. https://doi.org/10.1103/PhysRev.138.A763

Bachurin, D. V., Weygand, D., Gumbsch, P., 2010. Dislocation-grain boundary interaction in $<111>$ textured thin metal films. Acta materialia 58, 5232-5241. https://doi.org/10.1016/j.actamat.2010.05.037

Brailsford, A. D., 1966a. Stress field of a dislocation. Physical Review 142, 383-387. https://doi.org/10.1103/PhysRev.142.383

Brailsford, A. D., 1966b. Effective mass of a dislocation. Physical Review 142, 388-391.

https://doi.org/10.1103/PhysRev.142.388

Celli, V., Flytzanis, N., 1970. Motion of screw dislocation in a crystal. Journal of Applied Physics 41, 4443-4447. https://doi.org/10.1063/1.1658479

Chen, B., Lutker, K., Raju, S. V., Yan, J., Kanitpanyacharoen, W., Lei, J., Yang, S., Wenk, H.-R., Mao, H., Williams, Q., 2012. Texture of nanocrystalline nickel: Probing the lower size limit of dislocation activity. Science 338, 1448-1451. https://doi.org/10.1126/science.1228211

Chen, X., Xiong, L., McDowell, D., Chen, Y., 2017. Effects of phonons on mobility of dislocations and dislocation arrays. Scripta Materialia 137, 22-26. https://doi.org/10.1016/j.scriptamat.2017.04.033

Cho, J., Molinari, J. –F., Anciaux, G., 2017. Mobility law of dislocations with several character angles and temperatures in FCC aluminum. International Journal of Plasticity 90, 66-75. http://dx.doi.org/10.1016/j.ijplas.2016.12.004

Du, H., Jia, C. –L., Houben, L., Metlenko, V., De Souza, R. A., Waser, R., Mayer, J., 2015. Atomic structure and chemistry of dislocation cores at low-angle tilt grain boundary in $SrTiO_3$ bicrystals. Acta Materialia 89, 344-351. https://doi.org/10.1016/j.actamat.2015.02.016

Eshelby, J. D., 1956. Supersonic dislocations and dislocations in dispersive media. Proceedings of the Physical Society B 69, 1013-1019. https://doi.org/10.1088/0370-1301/69/10/307

Eshelby, J. D., 1962. The interaction of kinks and elastic waves. Proceedings of the Royal Society A 266, 222-246. https://doi.org/10.1098/rspa.1962.0058

Filippova, V. P., Kunavin, S. A., Pugachev, M. S., 2014. Calculation of the parameters of the Lennard-Jones potential for pairs of identical atoms based on the properties of solid substances. Materialovedenie 6, 3-6. https://doi.org/10.1134/S2075113315010062

Gu, Y., Xiang, Y., Srolovitz, D. J., El-Awady, J. A., 2018. Self-healing of low angle grain boundaries by vacancy diffusion and dislocation climb. Scripta Materialia 155. 155-159. https://doi.org/10.1016/j.scriptamat.2018.06.035

Gumbsch, P., Gao, H., 1999. Dislocations faster than the speed of sound. Science 283, 965-968. https://doi.org/10.1126/science.283.5404.965

Gurrutxaga-Lerma, B., 2016. The role of the mobility law of dislocations in the plastic response of shock loaded pure metals. Modelling and Simulation in Materials Science and Engineering. 24, 065006. https://doi.org/10.1088/0965-0393/24/6/065006

Heitler, W., 1944. The Quantum theory of Radiation, 2nd Ed., Oxford.

Hirth, J. P., Lothe. J., 1982. Theory of dislocations, Wiley.

Hirth, J. P., Zbib, H. M., Lothe, J., 1998. Forces on high velocity dislocations. Modelling and Simulation in Materials Science and Engineering 6, 165-169. https://doi.org/10.1088/0965-0393/6/2/006

Hoover, W. G., 1985. Canonical dynamics: equilibrium phase-space distributions. Physical Review A 31, 1695-1697. https://doi.org/10.1103/PhysRevA.31.1695

Hughes, D. A., Hansen, N., 2014. Exploring the limit of dislocation based plasticity in nanostructured metals. Physical Review Letters 112, 135504. https://doi.org/10.1103/PhysRevLett.112.135504

Hull, D., Bacon, D. J., 2011. Introduction to dislocations, 5th Ed., Elsevier.

Kim, H. –S, Kang, S. D., Tang, Y., Hanus, R., Snyder, G. J., 2016. Dislocation strain as the mechanism of phonon scattering at grain boundaries. Materials Horizons 3, 234-240. https://doi.org/10.1039/c5mh00299k

Kim, S. I., Lee, K. H., Mun, H. A., Kim, H. S., Hwang, S. W., Roh, J. W., Yang, D. J., Shin. W. H., Li, X. S., Lee, Y. H., Snyder, G. J., Kim, S. W., 2015. Dense dislocation arrays embedded in grain boundaries for high-performance bulk thermoelectrics. Science 348, 109-114. https://doi.org/10.1126/science.aaa4166

Kim, S., Ho, D. T., Kang, K., Kim, S. Y., 2016. Phonon scattering during dislocation motion inducing stress-drop in cubic metals. Acta Materialia 115, 143-154. https://doi.org/10.1016/j.actamat.2016.05.053

Kim, S., Kim, H., Kang, K., Kim, S. Y., 2019. Relativistic effect inducing drag on fast-moving dislocation in discrete system. International Journal of Plasticity (accepted).

Koizumi, H., Kirchner, H. O. K., Suzuki, T., 2002. Lattice wave emission from a moving dislocation. Physical Review B 65, 214104. https://doi.org/10.1103/PhysRevB.65.214104

Krasnikov, V. S., Mayer, A. E., 2018. Influence of local stresses on motion of edge dislocation in aluminum. International Journal of Plasticity 101, 170-187. https://doi.org/10.1016/j.ijplas.2017.11.002

Kresse, O., Truskinovsky, L., 2003. Mobility of lattice defects: discrete and continuum approaches. Journal of the Mechanics and Physics of Solids 51, 1305-1332. https://doi.org/10.1016/S0022-5096(03)00019-X

Kresse, O., Truskinovsky, L., 2004. Lattice friction for crystalline defects: from dislocations to cracks. Journal of the Mechanics and Physics of Solids 52, 2521-2543. https://doi.org/10.1016/j.jmps.2004.04.011

Li, R., Chew, B. W., 2017. Grain boundary traction signatures: Quantifying the asymmetrical dislocation emission processes under tension and compression. Journal of the Mechanics and Physics of Solids 103, 142-154. https://doi.org/10.1016/j.jmps.2017.03.009

Li, X., Wei. Y., Lu, L., Lu, K., Gao, H., 2010. Dislocation nucleation governed softening and maximum strength in nano-twinned metals. Nature 464, 877-880. https://doi.org/10.1038/nature08929

Lifshitz, I. M., Kosevich, A. M., 1966. The dynamics of a crystal lattice with defects. Reports on Progress in Physics 29, 217-254. https://doi.org/10.1088/0034-4885/29/1/305

Lim, A. T., Haataja, M., Cai, W., Srolovitz, D. J., 2012. Stress-driven migration of simple low-angle mixed grain boundaries. Acta Materialia 60, 1395-1407. https://doi.org/10.1016/j.actamat.2011.11.032

Maradudin, A. A., Mazur, P., Montroll, E. W., Weiss, G. H., 1958. Remarks on the vibrations of diatomic lattices. Review of Modern Physcis 30, 175-196. https://doi.org/10.1103/RevModPhys.30.175

Markenscoff, X., Ni, L., 2001. The transient motion of a dislocation with a ramp-like core. Journal of the Mechanics and Physics of Solids 49, 1603-1619. https://doi.org/10.1016/S0022-5096(00)00062-4

Mendelev, M. I., Han, S., Srolovitz, D. J., Ackland, G. J., Sun, D. Y., Asta, M., 2003. Development of new interatomic potentials appropriate for crystalline and liquid iron.

Philosophical Magazine A 83, 3977-3994. https://doi.org/10.1080/14786430310001613264

Mills, M. J., Stadelmann, P., 1989. A study of the structure of Lomer and 60° dislocations in aluminium using high-resolution transmission electron microscopy. Philosophical Magazine A 60, 355-384. https://doi.org/10.1080/01418618908213867

Mishin, Y., Farkas, D., Mehl, M. J., Papaconstantopoulos, D. A., 1999. Interatomic potentials for monoatomic metals from experimental data and *ab initio* calculations. Physical Review B 59, 3393. https://doi.org/10.1103/PhysRevB.59.3393

Mishin, Y., Mehl, M. J., Papaconstantopoulos D. A., Voter, A. F., Kress, J. D., 2001. Structural stability and lattice defects in copper: *Ab initio*, tight-binding, and embedded-atom calculations. Physical Review B 63, 224106. https://doi.org/10.1103/PhysRevB.63.224106

Montroll, E. W., Potts, R. B., 1955. Effect of defects on lattice vibrations. Physical Review 100, 525-543. https://doi.org/10.1103/PhysRev.100.525

Nosé, S., 1984. A unified formulation of the constant temperature molecular dynamics methods. Journal of Chemical Physics 81, 511-519. https://doi.org/10.1063/1.447334

Ohashi, K., 1968. Scattering of lattice waves by dislocations. Journal of the Phyical Society of Japan 24, 437-445. https://doi.org/10.1143/JPSJ.24.437

Pellegrini, Y. –P., 2012. Screw and edge dislocations with time-dependent core width: from dynamical core equations to an equation of motion. Journal of the Mechanics and Physics of Solids 60, 227-249. https://doi.org/10.1016/j.jmps.2011.11.002

Pellegrini, Y. –P., 2014. Equation of motion and subsonic-transonic transitions of rectilinear edge dislocations: A collective-variable approach. Physical Review B 90, 054120. https://doi.org/10.1103/PhysRevB.90.054120

Peng, S., Wei, Y., Jin, Z., Yang, W., 2019. Supersonic screw dislocations gliding at the shear wave speed. Physical Review Letters 122, 045501. https://doi.org/10.1103/PhysRevLett.122.045501

Plimpton, S., 1995. Fast parallel algorithms for short-range molecular dynamics. Journal of Computational Physcis 117, 1-19. https://doi.org/10.1006/jcph.1995.1039

Quek, S., S., Chooi, Z. H., Wu, Z., Zhang, Y. W., Srolovitz, D., 2016. The inverse hall-petch relation in nanocrystalline metals: A discrete dislocation dynamics analysis. Journal of the Mechanics and Physics of Solids 88, 252-266. https://doi.org/10.1016/j.jmps.2015.12.012

Read, W. T., Shockley, W., 1950. Dislocation models of crystal grain boundaries. Physical Review 78, 275-289. https://doi.org/10.1103/PhysRev.78.275

Sakamoto, M., 1991. High-velocity dislocations: Effective mass, effective line tension and multiplication. Philosophical Magazine A 63, 1241-1248. https://doi.org/10.1080/01418619108205580

Sanosoz, F., Molinary, J. F., 2005. Mechanical behavior of $\Sigma$ tilt grain boundaries in nanoscale Cu and Al: A quasicontinuum study. Acta Materialia 53, 1931-1944. https://doi.org/10.1016/j.actamat.2005.01.007

Shan, Z., Stach, E. A., Wiezorek, J. M. K., Knapp, J. A., Follstaedt, D. M., Mao, S. X., 2004. Grain boundary-mediated plasticity in nanocrystalline nickel. Science 305, 654-657. https://doi.org/10.1126/science.1098741

Smirnova, D. E., Kuksin, A. Y., Starikov, S. V., Stegailov, V. V., Insepov, Z., Rest, J., Yacout, A. M., 2013. A ternary EAM interatomic potential for U-Mo alloys with xenon. Modelling and Simulation in Materials Science and Engineering 21, 035011. https://doi.org/10.1088/0965-0393/21/3/035011

Srinivasan, S. G., Liao, X. Z., Baskes, M. I., McCabe, R. J., Zhao, Y. H., Zhu, Y. T., 2005. Compact and dissociated dislocations in aluminum: Implications for deformation. Physical Review Letters 94, 125502. https://doi.org/10.1103/PhysRevLett.94.125502

Sutton, A. P., Balluffi, R. W., 1995. Interfaces in Crystalline Materials, Oxford University Press.

Swinburne, T. D., Dudarev, S. L., 2015. Phonon drag force acting on a mobile crystal defect: Full treatment of discreteness and nonlinearity. Physical Review B 92, 134302. https://doi.org/ 10.1103/PhysRevB.92.134302

Tsuzuki, H., Branicio, P. S., Rino, J. P., 2008. Accelerating dislocations to transonic and supersonic speeds in anisotropic metals. Applied Physics Letters 92, 191909. https://doi.org/10.1063/1.2921786

Verschueren, J., Gurrutxaga-Lerma, B., Balint, D. S., Sutton, A. P., Dini, D., 2018. Instabilities of high speed dislocations. Physical Review Letters 121, 145502. https://doi.org/10.1103/PhysRevLett.121.145502

Wang, L., Abeyaratne, R., 2018. A one-dimensional peridynamic model of defect propagation and its relation to certain other continuum models. Journal of the Mechanics and Physics of Solids 116, 334-349. https://doi.org/10.1016/j.jmps.2018.03.028

Wang, Z. Q., Beyerlein, I. J., 2008. Stress orientation and relativistic effects on the separation of moving screw dislocations. Physical Review B 77, 184112. https://doi.org/10.1103/PhysRevB.77.184112

Wei, Y., Peng, S. Y., 2017. The stress-velocity relationship of twinning partial dislocations and the phonon-based physical interpretation. Science China-Physics, Mechanics & Astoronomy 60, 114611. https://doi.org/10.1007/s11433-017-9076-8

Weinberger, C. R., 2010. Dislocation drag at the nanoscale. Acta Materialia 58, 6535-6541.

Winning, M., Rollett, A. D., Gottstein, G., Srolovitz, D. J., Lim, A., Shvindlerman, L. S., 2010. Mobility of low-angle grain boundaries in pure metals. Philosophical Magazine 90, 3107-3128. https://doi.org/10.1080/14786435.2010.481272

Yasaei, P., Fathizadeh, A., Hantehzadeh, R., Majee, A. K., El-Ghandour, A., Estrada, D., Foster, C., Aksamija, Z., Khalili-Araghi, F., Salehi-Khojin, A., 2015. Bimodal phonon scattering in graphene grain boundaries. Nano Letters 15, 4532-4540. https://doi.org/10.1021/acs.nanolett.5b01100

Zhang, L., Lu, C., Tieu, K., Shibuta, Y., 2018. Dynamic interaction between grain boundary and stacking fault tetrahedron. Scripta Materialia 144, 78-83. https://doi.org/10.1016/j.scriptamat.2017.09.027

**Figures and tables**

**Table 1.** Input parameters used to solve Eq. (21) for perfect and extended edge dislocations in cubic crystals at 0 K. Parameters in shaded cells were obtained phenomenologically by fitting the solutions of Eq. (21) to MD simulation results. Other parameters were derived by the linear elasticity theory or direct measurements of the MD simulation results.

| Material | $b\,[\text{Å}]$ | $C_t\,[km/s]$ | $w_0\,[\times 10^{13}\,s^{-1}]$ | $M_{dis}\,[km/(s \cdot GPa)]$ | $C_1^{dis}$ | $C_2^{dis}$ |
|---|---|---|---|---|---|---|
| Fe | 2.473 | 2.99 | 7.31 | 2.78 | 0.91 | 1.004 |
| Mo | 2.745 | 4.08 | 9.10 | 2.08 | 0.92 | 1.002 |
| Al | 2.864 | 3.22 | 6.89 | 6.13 | 0.96 | 1.04 |
| Cu | 2.556 | 2.15 | 5.07 | 3.67 | 0.96 | 1.03 |
| Ni | 2.489 | 2.89 | 7.00 | 3.78 | 0.97 | 1.02 |
| Au | 2.884 | 1.18 | 2.46 | 1.77 | 0.96 | 1.02 |

**Table 2.** Input parameters used to solve Eq. (21) for perfect and extended edge dislocations in cubic crystals at 300 K. Parameters in shaded cells were obtained phenomenologically by fitting the solutions of Eq. (21) to MD simulation results. Other parameters were derived by the linear elasticity theory or direct measurements of the MD simulation results. Here, $C_t (= \sqrt{G/\rho})$ is not included because the changes in shear modulus and density of the crystals are so small that there is no significant difference in $C_t$ compared to that at 0 K.

| Material | $w_0 [\times 10^{13} s^{-1}]$ | $M_{dis} [km/(s \cdot GPa)]$ | $C_1^{dis}$ | $C_2^{dis}$ |
|---|---|---|---|---|
| Fe | 6.94 | 4.65 | 1.45 | 1.35 |
| Mo | 8.70 | 3.00 | 1.55 | 1.30 |
| Al | 6.76 | 16.24 | 1.28 | 1.30 |
| Ni | 6.76 | 14.95 | 2.55 | 1.90 |

**Table 3.** Input parameters used to solve Eq. (26) for LAGBs at 0 K. Parameters in shaded cells were obtained phenomenologically by fitting the solutions of Eq. (26) to MD simulation results. Here, $w_0$ and $M_{dis}$ were obtained by MD simulations of perfect lattice and single edge dislocations by using an LJ potential with the same parameters, respectively. Since all the LAGBs used in this study consist of same edge dislocations, they share the same $M_{dis}$.

| $\sigma_{app}$ | $w_0 [\times 10^{13} s^{-1}]$ | $M_{dis} [m/(s \cdot GPa)]$ | $C_1^{LAGB}$ | $C_2^{LAGB}$ |
|---|---|---|---|---|
| 5.60 GPa | | | 85 | 0.175 |
| 6.02 GPa | 7.37 | 70 | 84 | 0.180 |
| 6.88 GPa | | | 83 | 0.185 |
| 7.30 GPa | | | 82 | 0.190 |

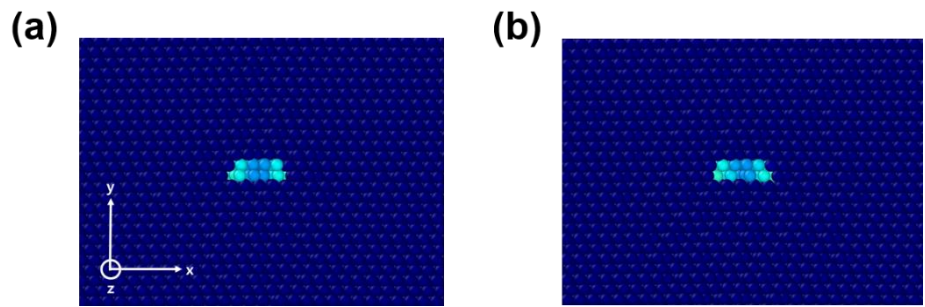

**Figure 1.** Equilibrium core structures of an edge dislocation in (a) iron and (b) molybdenum at 0 K. Color represents a centro-symmetry parameter.

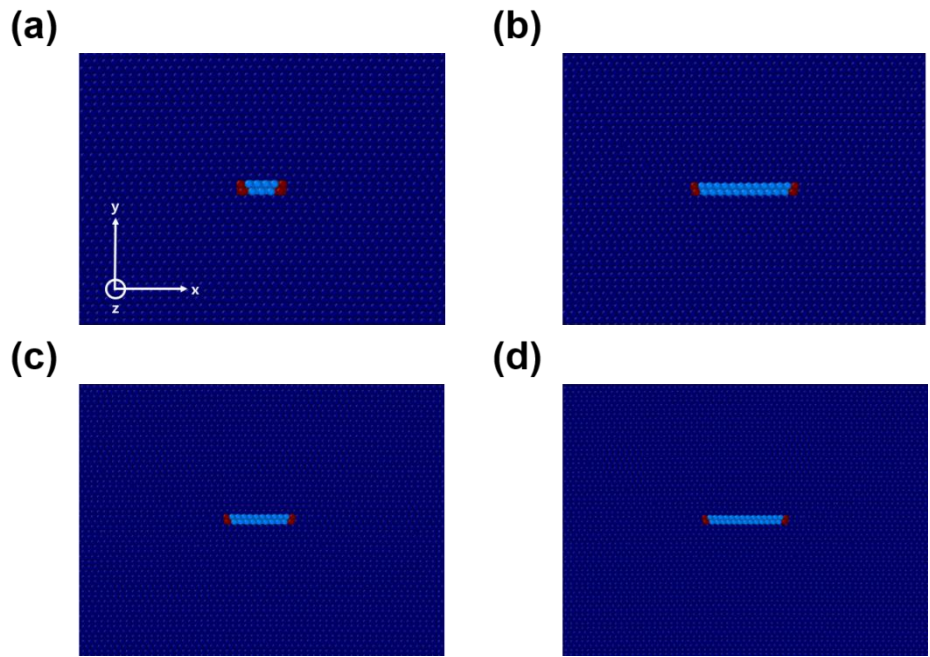

**Figure 2.** Equilibrium core structures of an extended edge dislocation in (a) aluminum, (b) copper, (c) nickel, and (d) gold at 0 K. Color represents a common neighbor analysis. Red atoms are partial dislocations, and light-blue atoms between red atoms are stacking faults.

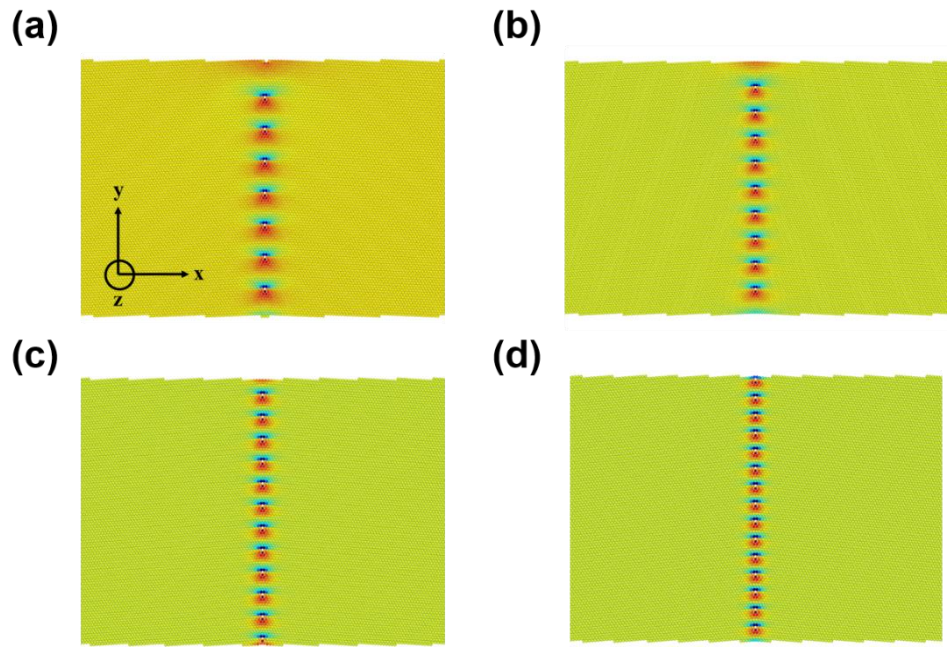

**Figure 3.** Equilibrium structures of symmetric tilt LAGBs at 0 K with misorientation angles of (a) $\theta = 5.09°$, (b) $\theta = 6.01°$, (c) $\theta = 7.34°$, and (d) $\theta = 9.42°$. Each LAGB consists of the same edge dislocations whose Burgers vector has only an $x$ component. Color represents $\sigma_{xx}$.

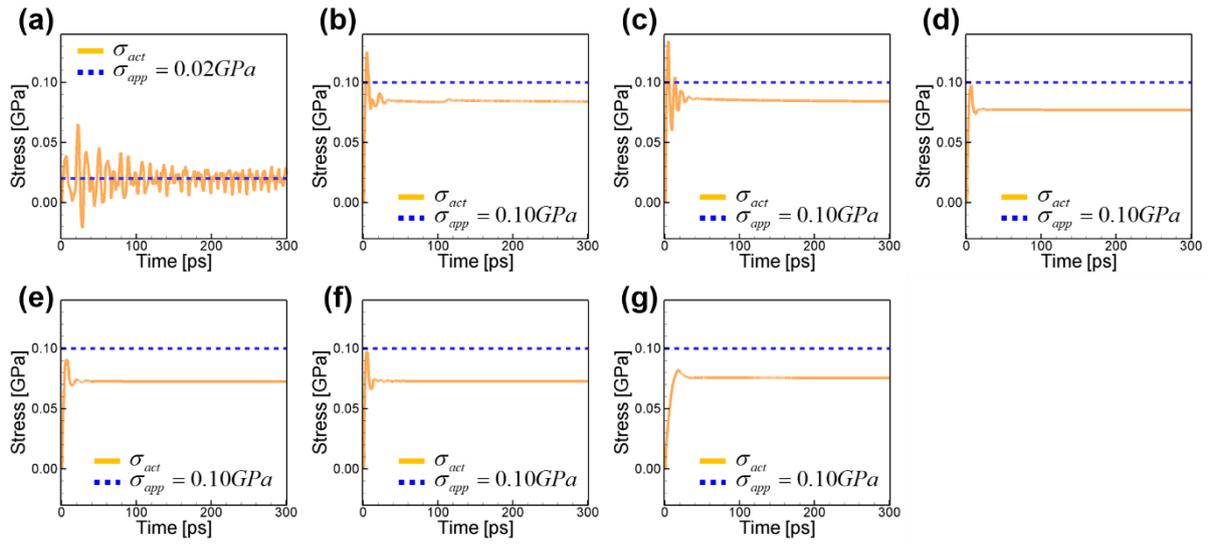

**Figure 4.** Variations in actual stress inside a system with a single edge dislocation over simulation time when an external stress, $\sigma_{app}$, is applied. Each figure corresponds to (a) iron when $\sigma_{app} < \sigma_P$ and (b) when $\sigma_{app} > \sigma_P$, (c) molybdenum when $\sigma_{app} > \sigma_P$, (d) aluminum when $\sigma_{app} > \sigma_P$, (e) copper when $\sigma_{app} > \sigma_P$, (f) nickel when $\sigma_{app} > \sigma_P$, and (g) gold when $\sigma_{app} > \sigma_P$.

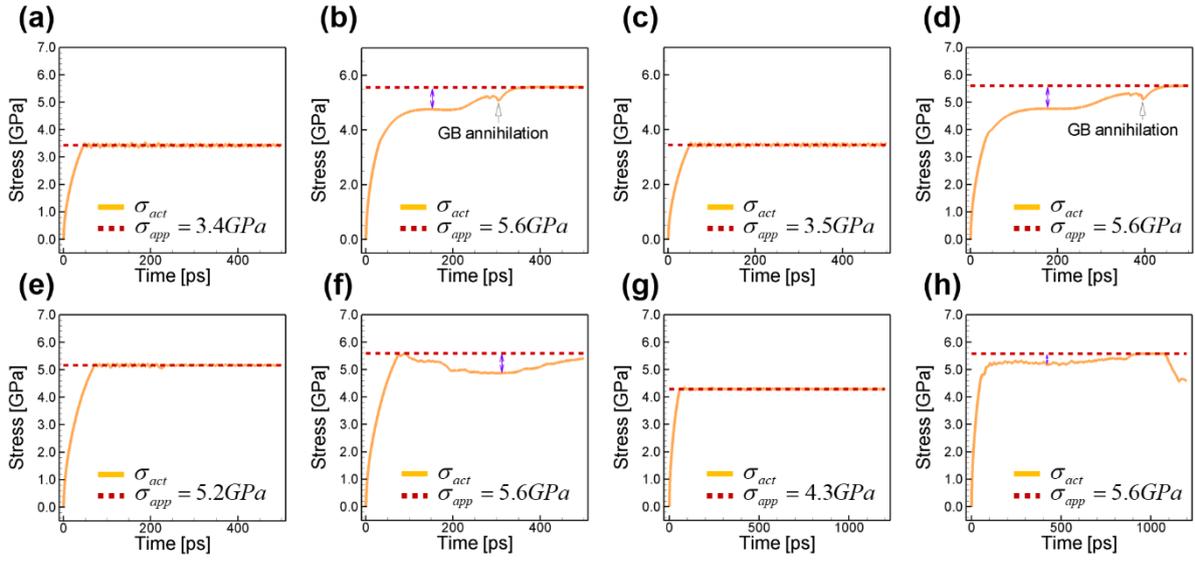

**Figure 5.** Variations in actual stress inside a system with an inserted LAGB over simulation time when an external stress, $\sigma_{app}$, is applied. Each figure corresponds to $\theta = 5.09°$ (a) when $\sigma_{app} < \sigma_P$ and (b) $\sigma_{app} > \sigma_P$, $\theta = 6.01°$ (c) when $\sigma_{app} < \sigma_P$ and (d) $\sigma_{app} > \sigma_P$, $\theta = 7.34°$ (e) when $\sigma_{app} < \sigma_P$ and (f) $\sigma_{app} > \sigma_P$, and $\theta = 9.42°$ (g) when $\sigma_{app} < \sigma_P$ and (h) $\sigma_{app} > \sigma_P$. Purple arrows in (b), (d), (f), and (h) represent the stress-drop while each LAGB is in a dynamic quasi-equilibrium state.

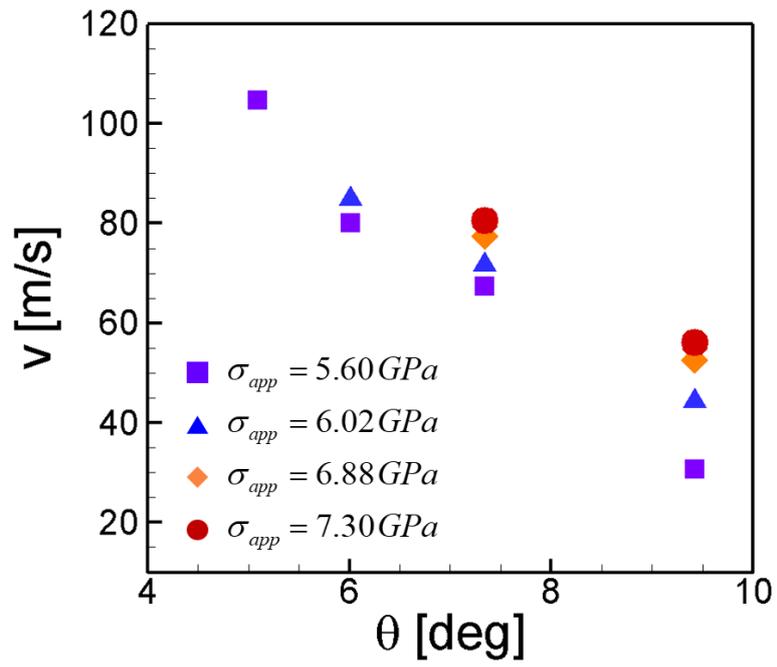

**Figure 6.** Relationship between misorientation angle and LAGB velocity.

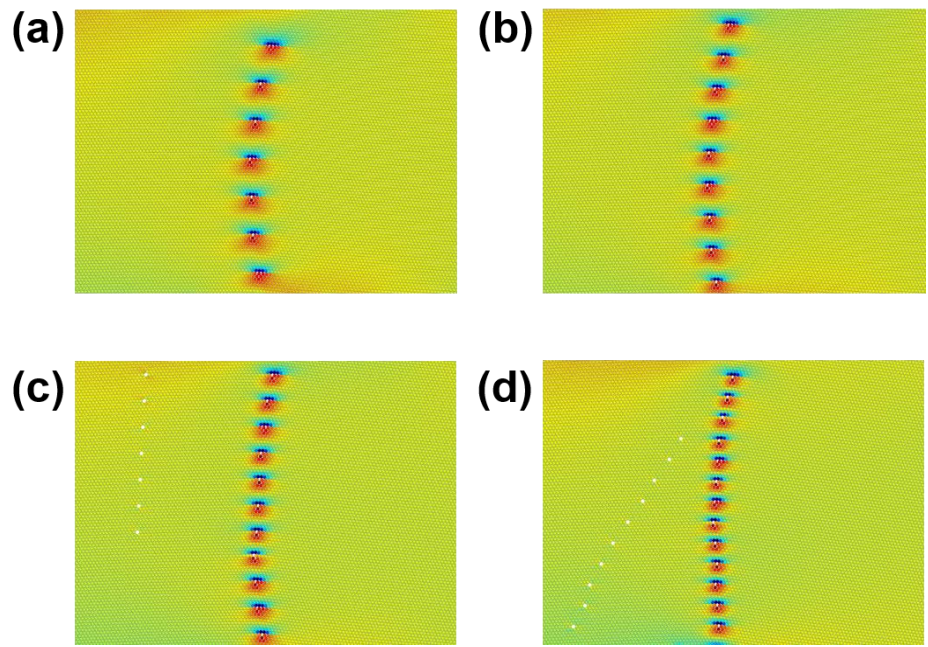

**Figure 7.** Curved shapes of LAGBs in motion by applied stress. Each figure corresponds to (a) $\theta = 5.09°$, (b) $\theta = 6.01°$, (c) $\theta = 7.34°$, and (d) $\theta = 9.42°$.

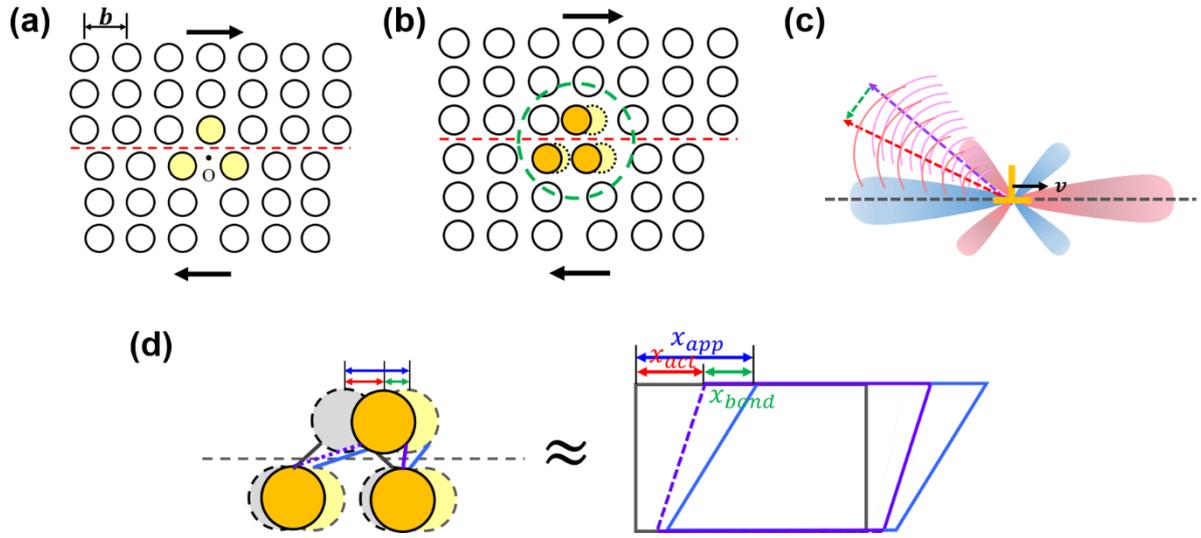

**Figure 8.** (a) Dislocation core when not moving, i.e., $\sigma_{app} < \sigma_P$. The core atoms are colored bright yellow. (b) Dislocation core when moving, i.e., $\sigma_{app} > \sigma_P$. As it moves, elastic waves emitted from the core are scattered and part of the energy dissipates from the core. As a result, the core retreats. The retreated core atoms are colored orange, and the scattering occurs within the region marked by the green dashed circle. (c) Schematic description of the scattering process. After the scattering by anharmonic strain field of the core, the initial radiated waves (marked by purple arrow) change to scattered waves (marked by red arrow) because of frequency shift and change of propagation path. As a result, the core moves backward, which is represented by green arrow. (d) Enlarged view of the region within the green circle in (b), where gray atoms represent initial core atoms without any loading. As $\sigma_{app}$ is applied, they deform to the bright yellow atoms. However, as phonon scattering, the core atoms retreat slightly. As a result, they finally deform to the orange-colored atoms. This process can roughly be described by simple shear deformation of the cell from a rectangular to parallelogram shape. Here, $x_{app}$ is the ideal elongation due to $\sigma_{app}$ if there is no scattering, and $x_{act}$ is the actual elongation resulting from phonon scattering.

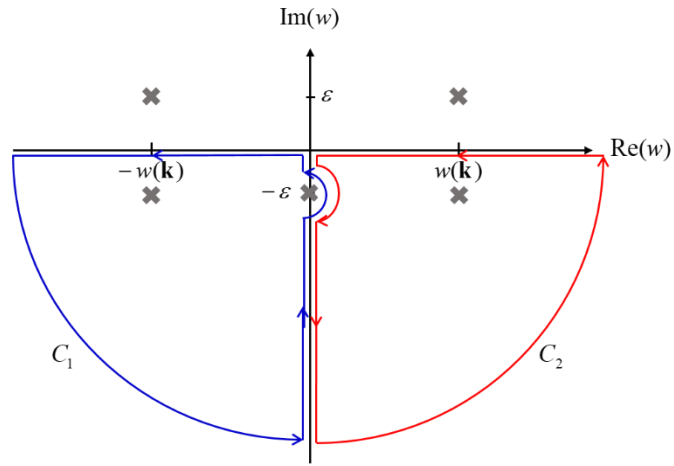

**Figure 9.** Integration path to calculate Eq. (7). Cross marks represent isolated poles, and blue and red paths represent contours $C_1$ and $C_2$, respectively.

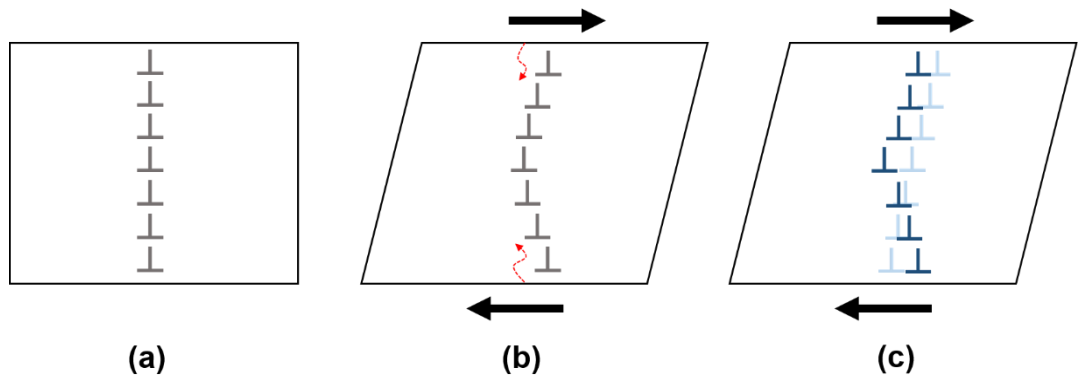

**Figure 10.** Graphical description of the structure of an LAGB at different times under shear stress. (a) Initial state when $\sigma_{app} = 0$. (b) Transition state of the LAGB under $\sigma_{app}\,(>\sigma_C)$, where $\sigma_C$ is the critical stress required to move the LAGB. Since the elastic wave supplied by $\sigma_{app}$ (marked by a red dashed arrow) travels from the free surface to the inside of the system, each dislocation comprising the LAGB begins its motion at a different time. This causes the structure of the LAGB to become curved. (c) The dynamic equilibrium state of the LAGB is marked by dark blue, whereas the sky blue-colored LAGB is the ideal structure expected when only linear elasticity theory is considered.

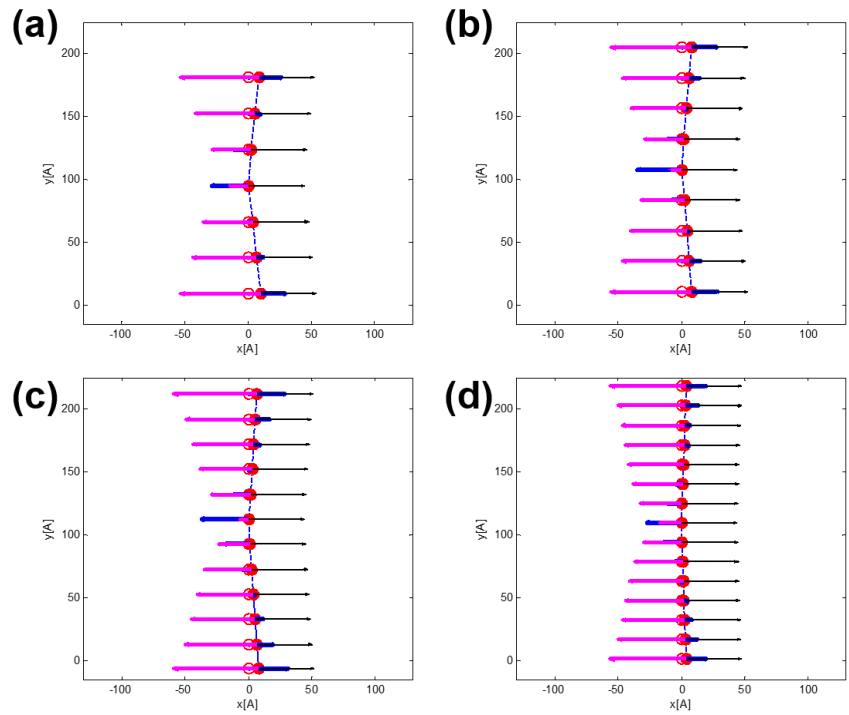

**Figure 11.** Force distributions acting on constituent dislocations of LAGBs for various misorientation angles. Hollow circles represent ideal structures of the LAGBs that are expected by only linear elasticity theory, and filled circles represent actual structures where the drag effect is considered. Black, blue, and purple arrows represent the external load, force from interactions among dislocations, and drag force acting on each dislocation, respectively. The length of the arrows is proportional to the magnitude of the force, and their directions correspond to those of the forces. Each figure describes the LAGB for which (a) $\theta = 5.09°$, (b) $\theta = 6.01°$, (c) $\theta = 7.34°$, and (d) $\theta = 9.42°$.

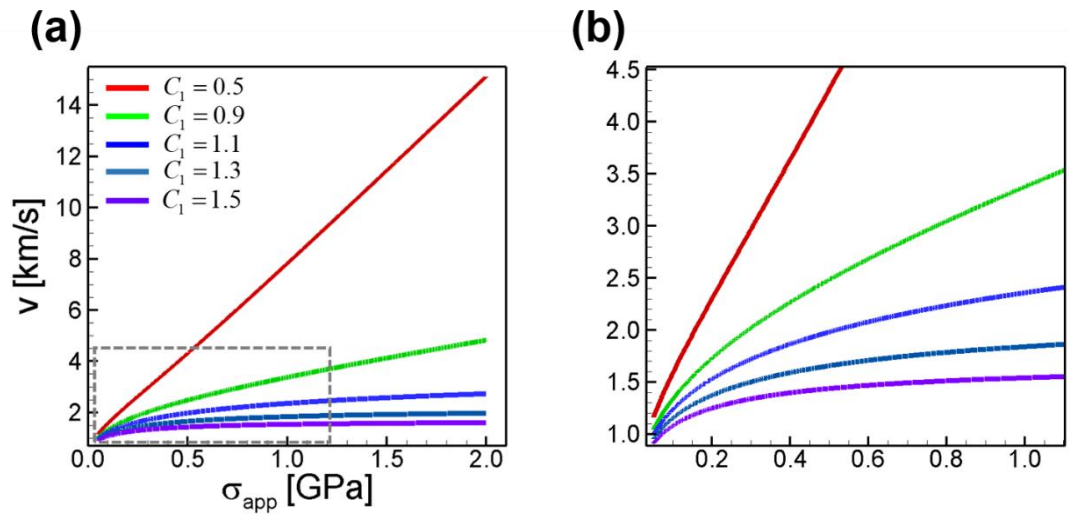

**Figure 12.** (a) Theoretical relationship between applied stress and dislocation speed obtained by solving Eq. (21) with changing $C_1^{dis}$. Here, we used the same input parameters used for nickel at 300 K, which are listed in Table 2, except with fixed $C_2^{dis} = 1.2$. (b) Enlarged graph of the gray rectangular area in (a).

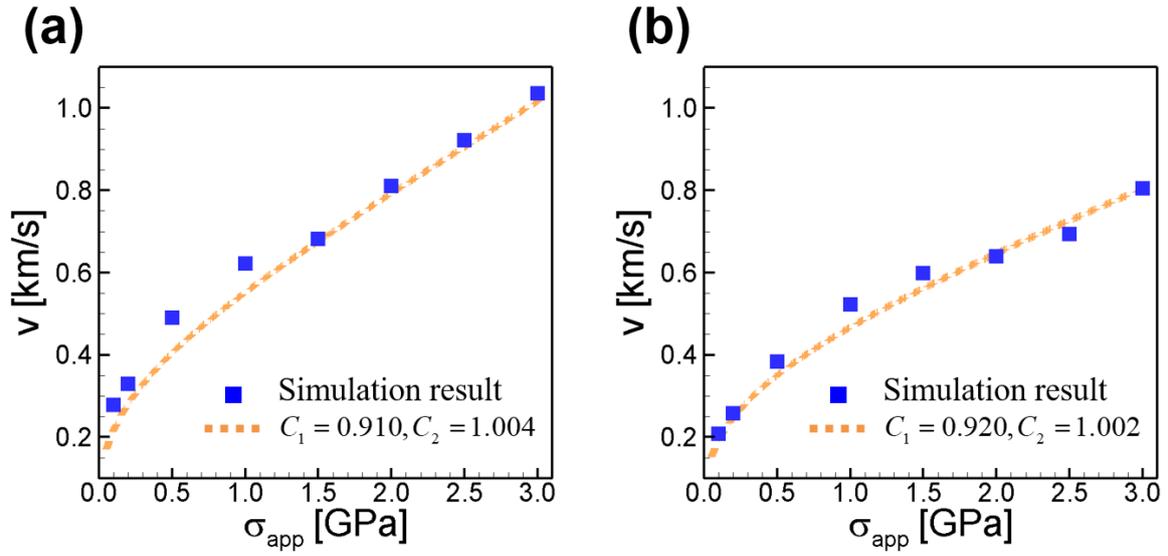

**Figure 13.** Relationship between applied stress and speed of a dislocation in (a) iron and (b) molybdenum at 0 K. Orange dashed lines correspond to solutions of Eq. (21) by inserting the input parameters listed in Table 1, and $C_1^{dis}$ and $C_2^{dis}$ are specified in each figure. Blue squares are results obtained by MD simulations.

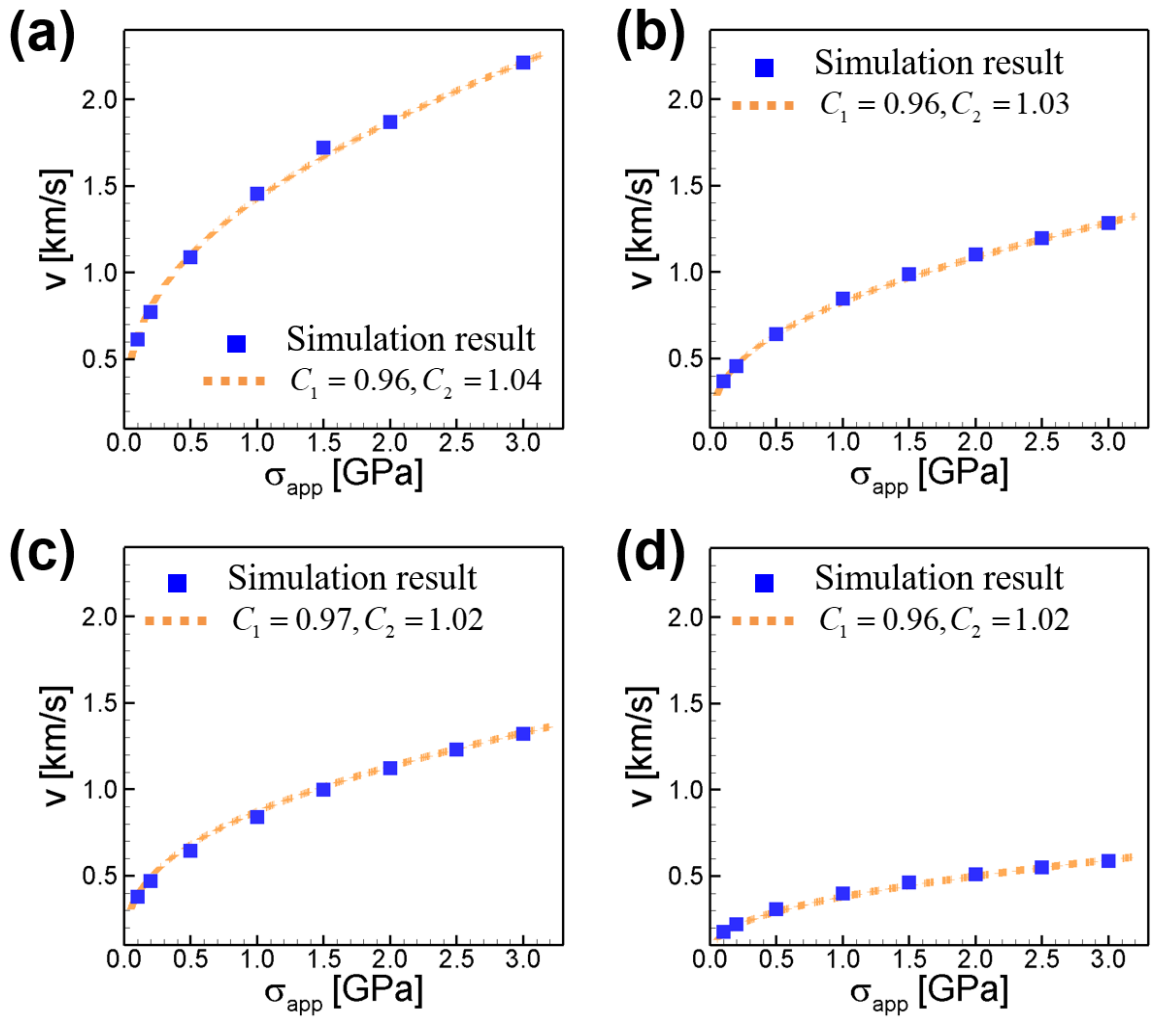

**Figure 14.** Relationship between applied stress and speed of an extended dislocation in (a) aluminum, (b) copper, (c) nickel, and (d) gold at 0 K. Orange dashed lines correspond to solutions of Eq. (21) obtained by inserting the input parameters listed in Table 1, and $C_1^{dis}$ and $C_2^{dis}$ are specified in each figure. Blue squares are results obtained by MD simulations.

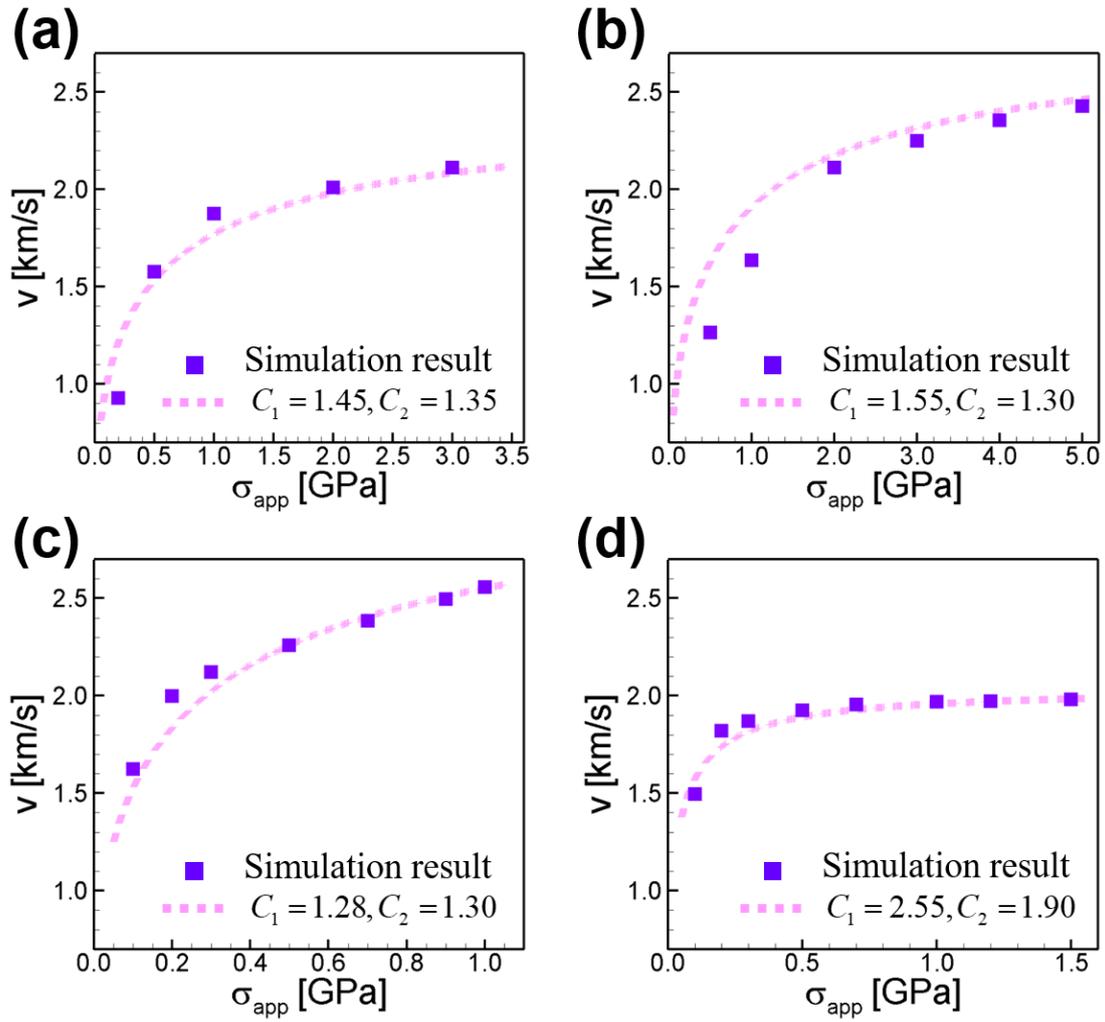

**Figure 15.** Relationship between applied stress and speed of an extended dislocation in (a) iron, (b) molybdenum, (c) aluminum, and (d) nickel at 300 K. Lavender dashed lines correspond to solutions of Eq. (21) obtained by inserting the input parameters listed in Table 2, and $C_1^{dis}$ and $C_2^{dis}$ are specified in each figure. Purple squares are results obtained by MD simulations.

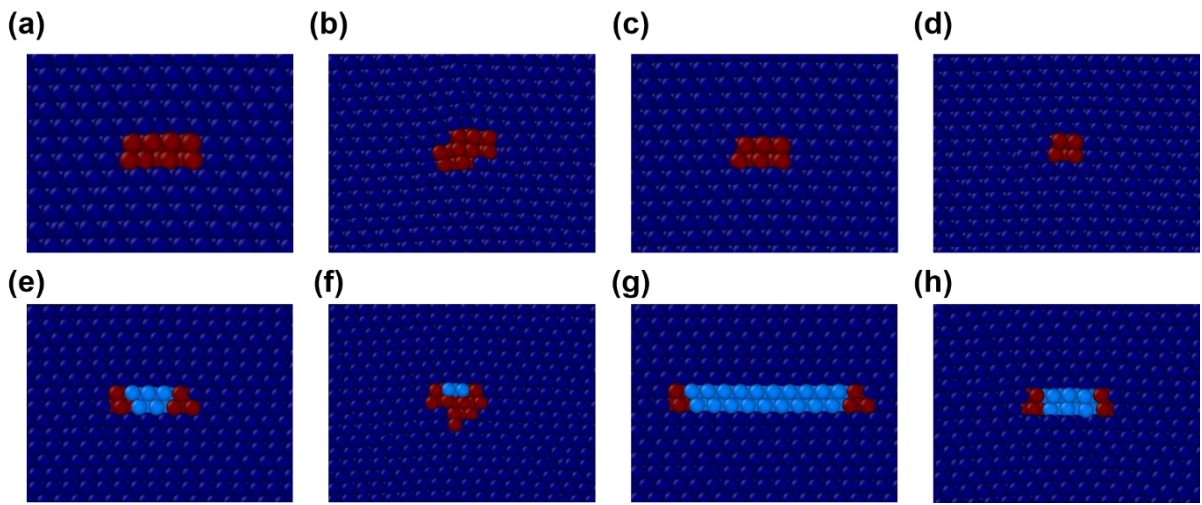

**Figure. 16.** Structures of dislocation cores while in motion at 0 K and 300 K. Edge dislocation core in iron at (a) 0 K and (b) 300 K under $\sigma_{app} = 3 GPa$. Edge dislocation core in molybdenum at (c) 0 K and (d) 300 K under $\sigma_{app} = 3 GPa$. Extended edge dislocation core in aluminum at (e) 0 K and (f) 300 K under $\sigma_{app} = 1 GPa$. Extended edge dislocation core in nickel at (g) 0 K and (h) 300 K under $\sigma_{app} = 1.5\ GPa$. Color represents a common neighbor analysis.

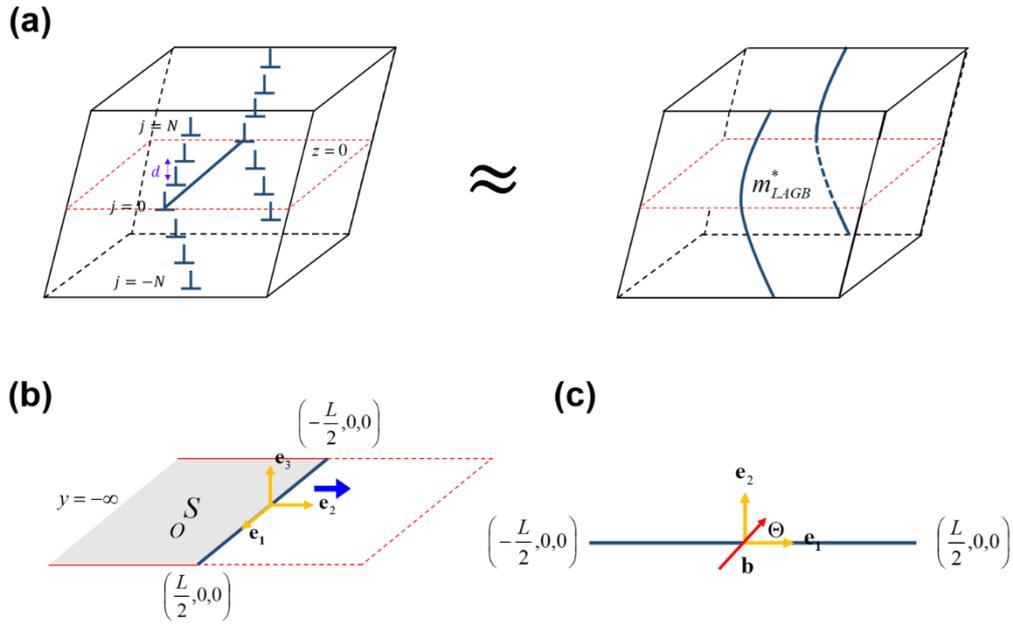

**Figure 17.** (a) Theoretical model of a curved LAGB under external stress. The LAGB consists of $2N+1$ edge dislocations, where the $0^{th}$ dislocation corresponds to the center of the LAGB. The right figure is an alternative continuum model where the array of dislocations is replaced with an oscillating string whose effective mass per thickness is defined as $m_{LAGB}^*$. (b) Enlarged configuration of the $z=0$ plane in (a). Note that we used a different coordinate system here than that defined in Figure 3. The shaded region represents an area slipped by the moving dislocation toward the positive $y$ direction, and the glide of the dislocation is confined to the $xy$ plane. (c) Enlarged configuration of the $0^{th}$ dislocation in the $z=0$ plane in (b). Here, $\Theta$ is the angle between the Burgers vector and the $x$ axis, and $L$ is the length of the dislocation.

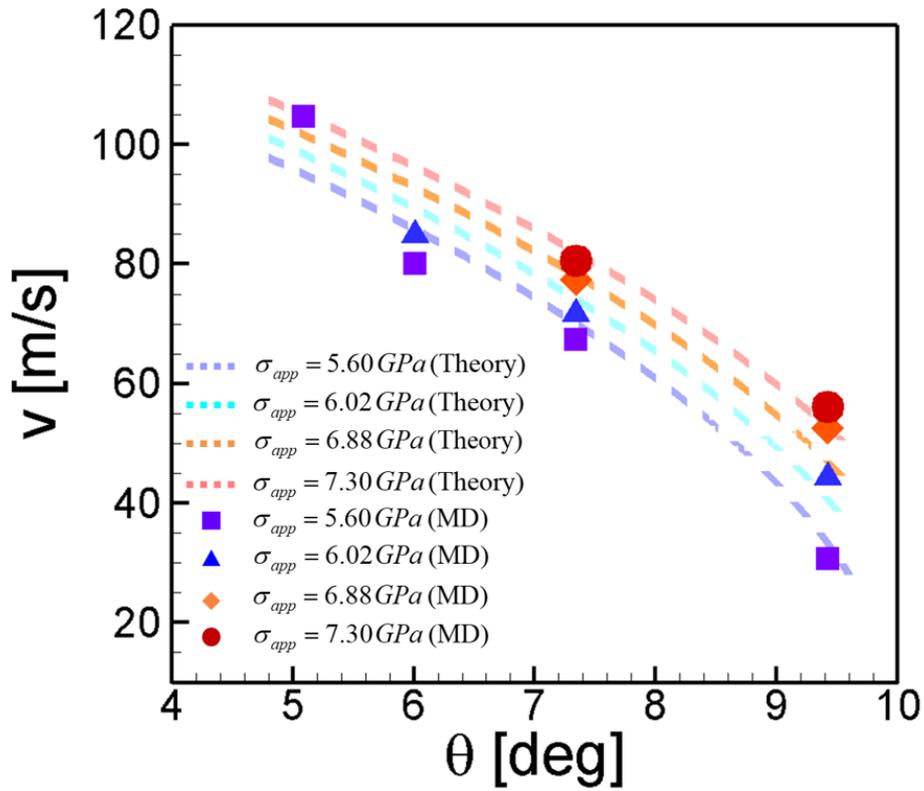

**Figure 18.** Relationship between misorientation angle and LAGB speed. Markers are the results of MD simulations, and dashed lines correspond to theoretical solutions of Eq. (26) under different stresses. The input parameters used to solve Eq. (26) are summarized in Table 3.